\documentclass[preprint,prd,aps,showpacs,showkeys,nofootinbib]{revtex4}
\usepackage{graphicx}
\usepackage{dcolumn}
\usepackage{bm}
\usepackage{color}
\usepackage[dvipsnames]{xcolor}
\usepackage{amssymb}

\definecolor{light-gray}{gray}{0.78}
\definecolor{mid-gray}{gray}{0.55}
\definecolor{dark-gray}{gray}{0.32}

\begin{document}
\title{Higgs boson decays $h\rightarrow Z \gamma$ and $h\rightarrow m_V Z$ in the $U(1)_X$VLFM}
\author{Rong-Zhi Sun$^{1,2,3}$, Shu-Min Zhao$^{1,2,3}$\footnote{zhaosm@hbu.edu.cn}, Yue-Tong Liu$^{1,2,3}$, Xing-Xing Dong$^{1,2,3,4}$\footnote{dongxx@hbu.edu.cn}}
\affiliation{$^1$ Department of Physics, Hebei University, Baoding 071002, China}
\affiliation{$^2$ Hebei Key Laboratory of High-precision Computation and Application of Quantum Field Theory, Baoding, 071002, China}
\affiliation{$^3$ Hebei Research Center of the Basic Discipline for Computational Physics, Baoding, 071002, China}
\affiliation{$^4$ Departamento de Fisica and CFTP, Instituto Superior T$\acute{e}$cnico, Universidade de Lisboa,
Av.Rovisco Pais 1,1049-001 Lisboa, Portugal}
\date{\today}

\begin{abstract}
We study the Higgs boson decays $h \to Z\gamma$ and $h \to m_VZ$ in a model with vectorlike fermions and $U(1)_X$ symmetry ($U(1)_X$VLFM), where $m_V$ is a vector meson ($\rho,\ \omega,\ \phi,\ J/\psi,\ \Upsilon$). The exotic Yukawa interactions in this model generate mixing between Standard Model (SM) fermions and vectorlike fermions, and this mixing affects the Higgs boson mass and Higgs couplings. The corrections to the CP-even and CP-odd $h\gamma Z$ couplings come from loop diagrams that involve the new particles, and these corrections have a clear impact on the decay rates of $h\to Z\gamma$ and $h\to m_VZ$. In suitable regions of the parameter space, the model can produce non-negligible deviations in $\Gamma_{\rm NP}(h\to Z\gamma)/\Gamma_{\rm SM}(h\to Z\gamma)$ and $\Gamma_{\rm NP}(h\to m_VZ)/\Gamma_{\rm SM}(h\to m_V Z)$, providing possible signals of new physics (NP) beyond the SM.
\end{abstract}

\maketitle
\section{introduction}
In 2012, the ATLAS and CMS collaborations at the Large Hadron Collider (LHC) discovered a Higgs boson with a mass of about 125 GeV~\cite{ATLAS:2012yve, CMS:2012qbp}, clearly confirming that Electro-Weak (EW) Symmetry Breaking (EWSB) is realized through the Higgs mechanism. Within the current theoretical and experimental precision, its properties are consistent with the predictions of the Standard Model (SM). Although the SM has achieved great success in describing known particles and interactions, it still fails to address several important issues. In particular, if fermions obtain their masses solely through Yukawa interactions with the Higgs field, a pronounced hierarchy among fermion masses arises. For example, the mass ratio of the top quark to the electron reaches $m_t/m_e \approx 3.5\times10^{5}$, while neutrino masses $m_\nu$ lie only at the eV scale~\cite{T2K:2011ypd, DayaBay:2012fng}, leading to $m_t/m_\nu \approx 1.4\times10^{12}$~\cite{PDG}. Furthermore, the SM contains no viable dark matter candidate, cannot explain the matter-antimatter asymmetry of the Universe, and does not resolve the hierarchy problem between the EW and Planck scales. These longstanding shortcomings motivate the exploration of possible extensions to the SM.

Therefore, a variety of representative theoretical frameworks beyond the SM have been proposed, including the Two-Higgs-Doublet Model (2HDM)~\cite{2HDM1,2HDM2}, the Minimal Supersymmetric Standard Model (MSSM) and its extensions such as the NMSSM \cite{Maniatis:2009re,Martin:1997ns}, as well as composite Higgs models in which the Higgs boson emerges as a composite pseudo-Nambu-Goldstone boson~\cite{Panico:2015jxa}. These extensions typically introduce additional scalar or fermionic degrees of freedom, making the Higgs interaction structure considerably richer than that of the SM. In many such models, new particles and interactions can induce flavor-changing or CP-violating Higgs couplings at tree level or loop level~\cite{2HDM1,MSSM,SUSY}, thereby modifying the standard Higgs couplings and generating non-standard effective Higgs vertices. Specifically, although the SM forbids the $h\gamma Z$ vertex at tree level, it can be generated by loop contributions from new particles~\cite{vertex1,vertex2}, thus becoming an important window for exploring new physics (NP).

In the SM, the decay of the Higgs boson into a $Z$ boson and a photon ($h\to Z\gamma$) occurs through a loop diagram process, with a predicted branching ratio of approximately BR($h\to Z\gamma$)=($1.5\pm0.1)\times 10^{-3}$ \cite{hZgSM1,hZgSM2}, which is similar in magnitude to $h\to\gamma\gamma$. Various extensions of the SM can alter this decay rate by introducing new particles into the loop diagram, making the ratio BR($h\to Z\gamma$)/BR($h\to\gamma\gamma$) a sensitive probe for detecting NP effects. Additionally, the observation of $h\to Z\gamma$ can further confirm the coupling structure between the Higgs boson and the EW gauge bosons, thereby deepening our understanding of the EWSB mechanism. Notably, this decay mode is also sensitive to potential anomalous trilinear Higgs self-couplings \cite{Degrassi:2019yix}, and its precise measurement can help test the SM prediction of this fundamental quantity. Using proton-proton collision data at $\sqrt s$ = 13 TeV with an integrated luminosity of about 140 $\rm fb^{-1}$, the ATLAS and CMS collaborations have recently reported the first evidence for this decay, with a combined statistical significance of 3.4$\sigma$. The measured signal strength is ($2.2\pm0.7$) times the SM prediction, corresponding to a branching ratio of $(3.4\pm1.1)\times10^{-3}$, which is consistent with theoretical expectations within 1.9$\sigma$~\cite{hZgexp}.

The rare weak radiative Higgs decays $h \to m_V \gamma$ and $h \to m_V Z$, where $m_V$ denotes a meson, have been extensively investigated in the literature~\cite{htomgamma1,htomgamma2,htomgamma3,htomz}. Since the photon carries only transverse polarization, the decay $h \to m_V \gamma$ can produce only transversely polarized vector mesons. In contrast, in the $h \to m_V Z$ channel, the final state $Z$ boson can be either longitudinally or transversely polarized, allowing both pseudoscalar and vector mesons to be generated. Depending on the decay topologies, the amplitude receives two types of contributions: a direct contribution, in which the Higgs couples directly to the quarks forming the meson, and an indirect contribution, where an off-shell electroweak gauge boson transitions into the meson through local matrix elements~\cite{indirect1,indirect2}. These two components exhibit substantial interference in the $h \to m_V \gamma$ channel~\cite{htomgamma1,htomgamma2,htomgamma3}. For the $h \to m_V Z$ process, the indirect contribution induced by the effective $h\gamma Z$ vertex typically dominates over the direct one, particularly when $m_V$ is a light vector meson~\cite{htomz}. Furthermore, QCD factorization has been applied to refine the theoretical description of $h \to m_V Z$ decays~\cite{other1,other2,other3,other4}.

In recent years, experimental studies of Higgs boson decays into a $Z$ boson and a vector meson have made steady progress. Although no evidence for these rare decay modes has been observed so far, current LHC measurements have already imposed stringent constraints on several channels. Using 137~$\rm fb^{-1}$ of proton-proton collision data collected at $\sqrt{s}=13$ TeV, the CMS collaboration searched for a 125 GeV Higgs boson decaying into $Z\rho^{0}(770)$ and $Z\phi(1020)$. The resulting 95$\%$ confidence level (CL) upper limits on the branching ratios were found to be 1.04-1.31$\%$ for $Z\rho^{0}$ and 0.31-0.40$\%$ for $Z\phi$, corresponding to approximately 740-940 and 730-950 times their respective SM predictions~\cite{hZVexp1}. CMS also reported a search for the decay $h \to ZJ/\psi$ using the same dataset, obtaining an upper limit of $1.9\times10^{-3}$, about 800 times the SM expectation~\cite{hZVexp2}. While the current sensitivity remains far above the SM branching fractions, the experimental bounds on such rare Higgs decays continue to improve. With the forthcoming High-Luminosity LHC (HL-LHC), the prospects for probing these channels will be significantly enhanced, offering a promising opportunity to explore Higgs properties and potential NP effects.

Over the past decade, the absence of experimental evidence for supersymmetric particles has shifted considerable attention toward non-supersymmetric extensions of the SM. Among these possibilities, $U(1)_X$ frameworks containing vectorlike fermions provide a particularly economical and predictive class of models, capable of influencing both EW observables and flavor physics. In the $U(1)_X$VLFM model considered in this work, one generation of vectorlike quarks, one generation of vectorlike leptons and two additional complex scalar fields are introduced, leading to a significantly richer phenomenology than that in the SM. A distinguishing feature of vectorlike fermions is that their left-handed and right-handed components carry identical SM gauge quantum numbers, allowing them to obtain gauge invariant masses without relying on EWSB~\cite{Aguilar-Saavedra:2013qpa}. Consequently, they do not generate sizable modifications to the Higgs production cross section, while still remaining accessible to direct searches at the LHC through their strong or EW production channels. Once vectorlike fermions mix with SM fermions, the couplings of the latter to the $W$, $Z$ and Higgs bosons deviate from their SM forms~\cite{Cao:2022mif}. This mixing not only violates the Glashow-Iliopoulos-Maiani (GIM) mechanism and induces tree-level flavor-changing neutral currents (FCNCs), but also introduces new sources of CP violation that can affect the electric dipole moments (EDMs) of leptons, quarks and neutrons~\cite{Cao:2023smj}. Furthermore, the interplay between vectorlike fermions and the $U(1)_X$ gauge symmetry naturally breaks lepton flavor universality (LFU), offering a potential explanation for the $b\to s$ anomalies and providing a framework that can be extended to the generation of neutrino masses. These features collectively highlight the theoretical and phenomenological significance of the $U(1)_X$VLFM model. Motivated by these considerations, we investigate the impact of this model on rare Higgs decay channels, focusing on $h\to Z\gamma$ and $h\to m_V Z$ with $m_V = \rho,\omega,\phi,J/\psi,\Upsilon$. Complementary analyses of $h\to \gamma\gamma$ and $h\to VV^*~(V = Z, W)$ are also presented. We derive the relevant Feynman rules and amplitudes, perform numerical parameter scans and identify regions of parameter space consistent with current experimental constraints.

The paper is organized as follows. In Sec.~\ref{sec2}, we briefly introduce the main content of the $U(1)_X$VLFM. In Sec.~\ref{sec3}, we present the analytical expressions of the Higgs boson decays $h\to Z\gamma$ and $h\to m_V Z$. The input parameters and numerical results are shown in Sec.~\ref{sec4}. Our discussion and conclusion are given in Sec.~\ref{sec5}. Finally, some mass matrices and couplings are collected in Appendix~\ref{A1}.

\section{the $U(1)_X$VLFM}\label{sec2}
The gauge group of the $U(1)_X$VLFM is $SU(3)_C \otimes SU(2)_L \otimes U(1)_Y \otimes U(1)_X$, and the local gauge symmetries are broken through the Higgs mechanism. Compared with the SM, the model introduces three generations of right-handed neutrinos $\nu_R$, two singlet Higgs fields $\phi$ and $S$, as well as one generation of vectorlike quarks, vectorlike leptons and vectorlike neutrino. The light neutrino masses are generated at the tree level via the seesaw mechanism. The neutral CP-even components of the three scalar fields $H$, $\phi$ and $S$ mix with each other, resulting in a $3\times 3$ mass squared matrix. To obtain the 125 GeV Higgs boson mass, loop corrections should be taken into account. The $U(1)_X$ charges of all SM fields are assigned zero. The new fields beyond the SM are listed in Table~\ref{superfields}.
\begin{table}
\caption{ The superfields in $U(1)_X$VLFM}
\begin{tabular}{|c|c|c|c|c|}
\hline
Superfields & $SU(3)_C$ & $SU(2)_L$ & $U(1)_Y$ & $U(1)_X$ \\
\hline
$\phi$ & 1 & 1 & 0 & $Q_a+Q_b$ \\
\hline
$S$ & 1 & 1 & 0 & $Q_a$ \\
\hline
$\nu_R$ & 1 & 1 & 0 & 0  \\
\hline
$d_{XL}$ & 3 & 1 & -1/3 & $Q_a$  \\
\hline
$u_{XL}$ & 3 & 1 & 2/3 & -$Q_a$  \\
\hline
$d_{XR}$ & $\bar{3}$ & 1 & 1/3 & $Q_b$ \\
\hline
$u_{XR}$ & $\bar{3}$ & 1 & -2/3 & -$Q_b$\\
\hline
$e_{XL}$ & 1 & 1 & -1 & $Q_a$ \\
\hline
$\nu_{XL}$ & 1 & 1 & 0 & -$Q_a$ \\
\hline
$e_{XR}$ & 1 & 1 & 1 & $Q_b$\\
\hline
$\nu_{XR}$ & 1 & 1 & 0 & -$Q_b$ \\
\hline
\end{tabular}
\label{superfields}
\end{table}

There are one Higgs doublet and two Higgs singlets, whose specific forms are as follows:
\begin{eqnarray}
&&H=\left(\begin{array}{c}H^0\\H^-\end{array}\right),
~~~~~~
H^0={1\over\sqrt{2}}\Big(v+\phi_H+i\sigma_H\Big)\label{su2D},
\\
&&\phi={1\over\sqrt{2}}\Big(v_P+\phi_P+i\sigma_P\Big),~~~~~~
S={1\over\sqrt{2}}\Big(v_S+\phi_S+i\sigma_S\Big)\label{su2S}.
\end{eqnarray}
In Eqs.(\ref{su2D}-\ref{su2S}), $v$, $v_P$ and $v_S$ denote the nonzero vacuum expectation values (VEVs) corresponding to the Higgs superfields $H$, $\phi$ and $S$, respectively.

The relevant Lagrangian of the $U(1)_X$VLFM reads as
\begin{eqnarray}
&&\mathcal{L}=-\mu^2_H H^\dagger H-\mu^2_P |\phi|^2 -\mu^2_X |S|^2+\lambda_H(H^\dagger H)^2+\lambda_P |\phi|^4 +\lambda_X |S|^4\nonumber\\&&
~~+\lambda_{HP}(H^\dagger H)|\phi|^2+\lambda_{HX}(H^\dagger H)|S|^2+\lambda_{PX}|S|^2|\phi|^2\nonumber\\&&
~~-S d^*_{XL,k}Y^*_{XD,jk}d_{R,j}-S u^*_{R,j}Y_{XU,jk}u_{XL,k}-S e^*_{XL,k}Y^*_{XE,jk}e_{R,j}\nonumber\\&&
~~-S \nu^*_{R,j}Y_{XN,jk}\nu_{XL,k}-h.c.\nonumber\\&&
~~-\phi d^*_{XL,k}Y^*_{PD,jk}d_{XR,j}-\phi u^*_{XR,j}Y_{PU,jk}u_{XL,k}-\phi e^*_{XL,k}Y^*_{PE,jk}e_{XR,j}\nonumber\\&&
~~-\phi \nu^*_{XR,j}Y_{PN,jk}\nu_{XL,k}-h.c.\nonumber\\&&
~~-Y^*_{u,jk}\bar q_{L,k}H u_{R,j}+Y^*_{d,jk}\bar q_{L,k}\tilde{H} d_{R,j}+Y^*_{e,jk}\bar l_{k}\tilde{H} e_{R,j}+h.c.
\end{eqnarray}

We denote the $U(1)_Y$ charge by $Y^Y$ and the $U(1)_X$ charge by $Y^X$. As discussed in the textbook~\cite{Peskin}, the SM is anomaly free. For the $U(1)_X$VLFM model considered here, the cancellation of gauge and gravitational anomalies can be summarized as follows:

1. The anomalies involving three $SU(2)_L$ gauge bosons vanish exactly as in the SM, and the same applies to the corresponding $SU(3)_C$ anomaly.

2. The anomalies with one $SU(3)_C$ or one $SU(2)_L$ gauge boson are proportional to $\mathrm{Tr}[t^a]=0$ or $\mathrm{Tr}[\tau^a]=0$, respectively, and therefore vanish.
	
3. The mixed anomalies involving one $U(1)_Y$ or $U(1)_X$ and two $SU(3)_C$ gauge bosons are proportional to
\begin{eqnarray}
&&\mathrm{Tr}[t^a t^b Y^{Y}] = \frac{1}{2}\delta^{ab}\sum_q Y^{Y}_q,\quad
\mathrm{Tr}[t^a t^b Y^{X}] = \frac{1}{2}\delta^{ab}\sum_q Y^{X}_q.
\end{eqnarray}

4. Similarly, the anomalies containing one $U(1)_Y$ or $U(1)_X$ boson and two $SU(2)_L$ bosons are proportional to
\begin{eqnarray}
&&
\mathrm{Tr}[\tau^a \tau^b Y^{Y}] = \frac{1}{2}\delta^{ab}\sum_L Y^{Y}_L,\quad
\mathrm{Tr}[\tau^a \tau^b Y^{X}] = \frac{1}{2}\delta^{ab}\sum_L Y^{X}_L.
\end{eqnarray}
	
5. The anomalies of the three $U(1)$ gauge bosons are classified into four types
\begin{eqnarray}
&&\mathrm{Tr}[Y^YY^YY^Y]=\sum_n(Y^Y_n)^3,\quad
\mathrm{Tr}[Y^XY^XY^X]=\sum_n(Y^X_n)^3,\nonumber\\&&
\mathrm{Tr}[Y^XY^YY^Y]=\sum_nY^X_n(Y^Y_n)^2,\quad
\mathrm{Tr}[Y^YY^XY^X]=\sum_nY^Y_n(Y^X_n)^2.
\end{eqnarray}

6. The gravitational anomaly with one $U(1)_Y$ or $U(1)_X$ gauge boson is proportional to
\begin{eqnarray}
&&\mathrm{Tr}[Y^Y]=\sum_n Y_n^Y, \quad
\mathrm{Tr}[Y^X]=\sum_n Y_n^X.
\end{eqnarray}

For the parts that do not involve $U(1)_X$, the anomaly conditions are identical to those in the SM and can be easily verified to vanish. The cancellation of the $U(1)_X$ anomalies is also ensured, even though their structure is considerably more intricate than that of the SM. Therefore, the $U(1)_X$VLFM model is anomaly free.

In the $U(1)_X$VLFM model, the coexistence of the two Abelian gauge groups $U(1)_Y$ and $U(1)_X$ leads to a new effect absent in the SM: gauge kinetic mixing. This effect can also be generated radiatively via the RGEs, even if it is set to zero at the $M_{GUT}$.

The covariant derivative of this model can be written in the general form
\begin{eqnarray}
&&D_\mu=\partial_\mu-i\left(\begin{array}{cc}Y^Y,&Y^X\end{array}\right)
\left(\begin{array}{cc}g_{Y},&g{'}_{{YX}}\\g{'}_{{XY}},&g{'}_{{X}}\end{array}\right)
\left(\begin{array}{c}A_{\mu}^{\prime Y} \\ A_{\mu}^{\prime X}\end{array}\right)\;,
\label{gauge1}
\end{eqnarray}
where $A_{\mu}^{\prime Y}$ and $A_{\mu}^{\prime X}$ denote the gauge fields of $U(1)_Y$ and $U(1)_X$, respectively. Since both Abelian gauge groups remain unbroken, one can perform a basis rotation using an orthogonal matrix $R$ ($R^T R=1$), yielding
\begin{eqnarray}
&&\left(\begin{array}{cc}g_{Y},&g{'}_{{YX}}\\g{'}_{{XY}},&g{'}_{{X}}\end{array}\right)
R^T=\left(\begin{array}{cc}g_{1},&g_{{YX}}\\0,&g_{{X}}\end{array}\right)\;,
\label{gauge2}
\end{eqnarray}
which redefines the $U(1)$ gauge fields as
\begin{eqnarray}
&&R\left(\begin{array}{c}A_{\mu}^{\prime Y} \\ A_{\mu}^{\prime X}\end{array}\right)
=\left(\begin{array}{c}A_{\mu}^{Y} \\ A_{\mu}^{X}\end{array}\right)\;.
\label{gauge3}
\end{eqnarray}

$g_X$ denotes the gauge coupling constant associated with the $U(1)_X$ symmetry, while $g_{YX}$ characterizes the gauge kinetic mixing between the $U(1)_Y$ and $U(1)_X$ gauge groups. The neutral gauge bosons $A^Y_\mu$, $V^3_\mu$ and $A^X_\mu$ mix together at the tree level, leading to the mass matrix in the $(A^Y_\mu, V^3_\mu, A^X_\mu)$ basis
\begin{eqnarray}
&&\left(\begin{array}{*{20}{c}}
\frac{1}{4}g_{1}^2 v^2 &~~ -\frac{1}{4}g_{1}g_{2} v^2 & ~~\frac{1}{4}g_{1}g_{{YX}} v^2 \\
-\frac{1}{4}g_{1}g_{2} v^2 &~~ \frac{1}{4}g_{2}^2 v^2 & ~~-\frac{1}{4}g_{2}g_{{YX}} v^2\\
\frac{1}{4}g_{1}g_{{YX}} v^2 &~~ -\frac{1}{4}g_{2}g_{{YX}} v^2 &~~\frac{1}{4}g_{{YX}}^2 v^2+\frac{1}{4}g_{{X}}^2\xi^2
\end{array}\right)\label{gauge matrix}
\end{eqnarray}
with $\xi^2=4(Q_a + Q_b)^2 v^2_P+ 4Q^2_a v^2_S$.

To diagonalize the mass matrix in Eqs.(\ref{gauge matrix}), we use a unitary transformation involving two mixing angles $\theta_W$ and $\theta_W'$
\begin{eqnarray}
&&\left(\begin{array}{*{20}{c}}
\gamma_\mu\\ [6pt]
Z_\mu\\ [6pt]
Z'_\mu
\end{array}\right)=
\left(\begin{array}{*{20}{c}}
\cos\theta_{W} & \sin\theta_{W} & 0 \\ [6pt]
-\sin\theta_{W}\cos\theta_{W}' & \cos\theta_{W}\cos\theta_{W}' & \sin\theta_{W}'\\ [6pt]
\sin\theta_{W}\sin\theta_{W}' & -\cos\theta_{W}'\sin\theta_{W}' & \cos\theta_{W}'
\end{array}\right)
\left(\begin{array}{*{20}{c}}
A^Y_\mu\\ [6pt]
V^3_\mu\\ [6pt]
A^{X}_\mu
\end{array}\right).
\end{eqnarray}

The additional mixing angle $\theta_W'$ appears in the couplings involving $Z$ and $Z'$, and is given by
\begin{eqnarray}
\sin^2\theta_{W}'=\frac{1}{2}-\frac{(g_{{YX}}^2-g_{1}^2-g_{2}^2)v^2+
g_{X}^2\xi^2}{2\sqrt{(g_{{YX}}^2+g_{1}^2+g_{2}^2)^2v^4+2g_{X}^2(g_{{YX}}^2-g_{1}^2-g_{2}^2)v^2\xi^2+4g_{X}^4\xi^4}}.
\end{eqnarray}

The exact mass eigenvalues in Eqs.(\ref{gauge matrix}) are
\begin{eqnarray}
&&\qquad\;\quad\;m_\gamma^2=0,\nonumber\\
&&\qquad\;\quad\;m_{Z,{Z^{'}}}^2=\frac{1}{8}\Big((g_{1}^2+g_2^2+g_{YX}^2)v^2+g_{X}^2 \xi^2 \nonumber\\
&&\qquad\;\qquad\;\qquad\;\mp\sqrt{[(g_{1}^2+g_2^2+g_{YX}^2)v^2+g_{X}^2 \xi^2]^2-4(g^2_1+g^2_2)g^2_X v^2 \xi^2}\Big).
\end{eqnarray}

The simplified Higgs potential is given by
\begin{eqnarray}
&&V=\mu^2_H H^\dagger H+\mu^2_P |\phi|^2 +\mu^2_X |S|^2-\lambda_H(H^\dagger H)^2-\lambda_P |\phi|^4 -\lambda_X |S|^4\nonumber\\&&
~~~~~~~-\lambda_{HP}(H^\dagger H)|\phi|^2-\lambda_{HX}(H^\dagger H)|S|^2-\lambda_{PX}|S|^2|\phi|^2.
\end{eqnarray}

The VEVs of the Higgs fields should satisfy the following equations
\begin{eqnarray}
&&2\lambda_H v^2-2\mu^2_H +\lambda_{HP} v^2_P+\lambda_{HX} v^2_S=0, \\
&&2\lambda_X v^2_S-2\mu^2_X +\lambda_{HX} v^2+\lambda_{PX} v^2_P=0,\\
&&2\lambda_P v^2_P-2\mu^2_P +\lambda_{HP} v^2+\lambda_{PX} v^2_S=0.
\end{eqnarray}

In the $({\phi}_{H}, {\phi}_{S}, {\phi}_{P})$ basis, the CP-even Higgs mass squared matrix is
\begin{eqnarray}
m^2_{h} = \left(
\begin{array}{ccc}
m_{{\phi}_{H}{\phi}_{H}} &-\lambda_{HX}v v_S &-\lambda_{HP}v v_P \\
-\lambda_{HX}v v_S &m_{{\phi}_{S}{\phi}_{S}} &-\lambda_{PX}v_P v_S \\
-\lambda_{HP}v v_P &-\lambda_{PX}v_P v_S &m_{{\phi}_{P}{\phi}_{P}} \end{array}
\right),
\end{eqnarray}
\begin{eqnarray}
&&m_{\phi_{H}\phi_{H}}=\frac{1}{2}\Big(-6\lambda_H v^2-\lambda_{HP}v^2_P-\lambda_{HX}v^2_S\Big)+\mu^2_H,
\\&&m_{\phi_{S}\phi_{S}} = \frac{1}{2}\Big(-6\lambda_X v^2_S-\lambda_{HX}v^2-\lambda_{PX}v^2_P\Big)+\mu^2_X,
\\ &&m_{\phi_{P}\phi_{P}} = \frac{1}{2}\Big(-6\lambda_P v_P^2-\lambda_{HP}v^2-\lambda_{PX}v^2_S\Big)+\mu^2_P.
\end{eqnarray}
This matrix is diagonalized by $Z^H$
\begin{eqnarray}
Z^H m^2_{h} Z^{H,\dagger} = m^{dia}_{2,h},
 \end{eqnarray}
with
\begin{eqnarray}
\phi_H=\sum\limits_{j} Z^H_{j1} h_j, \quad
\phi_S=\sum\limits_{j} Z^H_{j2} h_j, \quad
\phi_P=\sum\limits_{j} Z^H_{j3} h_j.
\end{eqnarray}

The mass squared matrix for the CP-odd Higgs bosons in the basis $({\sigma}_{H}, {\sigma}_{S}, {\sigma}_{P})$ is diagonalized by $Z^A$ via the relation $Z^A m^2_{A_h} Z^{A,\dagger} = m^{dia}_{2,A_h}$
\begin{eqnarray}
m^2_{A_h} = \left(
\begin{array}{ccc}
m_{{\sigma}_{H}{\sigma}_{H}} &0 &0 \\
0 &m_{{\sigma}_{S}{\sigma}_{S}} &0 \\
0 &0 &m_{{\sigma}_{P}{\sigma}_{P}}\end{array}
\right),
\end{eqnarray}
\begin{eqnarray}
&&m_{\sigma_{H}\sigma_{H}}=\frac{1}{2}\Big(-2\lambda_H v^2-\lambda_{HP}v^2_P-\lambda_{HX}v^2_S\Big)+\mu^2_H,
\\&&m_{\sigma_{S}\sigma_{S}} = \frac{1}{2}\Big(-2\lambda_X v^2_S-\lambda_{HX}v^2-\lambda_{PX}v^2_P\Big)+\mu^2_X,
\\ &&m_{\sigma_{P}\sigma_{P}} = \frac{1}{2}\Big(-2\lambda_P v_P^2-\lambda_{HP}v^2-\lambda_{PX}v^2_S\Big)+\mu^2_P.
\end{eqnarray}
Here
\begin{eqnarray}
\sigma_H=\sum\limits_{j} Z^A_{j1} A_{h,j}, \quad
\sigma_S=\sum\limits_{j} Z^A_{j2} A_{h,j}, \quad
\sigma_P=\sum\limits_{j} Z^A_{j3} A_{h,j}.
\end{eqnarray}

The down-type quark mass matrix in the $(d_L, d_{XL})$ and $(d_R^*, d_{XR}^*)$ basis is given by
\begin{equation}
m_d = \left(
\begin{array}{cc}
\frac{1}{\sqrt{2}}v Y^T_d &0\\
\frac{1}{\sqrt{2}}v_S Y^T_{XD}  &\frac{1}{\sqrt{2}}v_P Y^T_{PD}\end{array}
\right),
\end{equation}
which is diagonalized by $U^d_L$ and $U^d_R$ according to
\begin{equation}
U_L^{d,*}\, m_d \, U_R^{d,\dagger} = m_d^{dia}.
\end{equation}

Furthermore, the mass matrices for the up-type quark and lepton are derived in the same way and listed in the Appendix~\ref{A1}.

We now introduce the couplings needed in this model, and begin by presenting some interactions involving the $Z$ boson. In the below equations, $P_L=\frac{1-{\gamma}_5}{2}$ and $P_R=\frac{1+{\gamma}_5}{2}$.
\begin{eqnarray}
&&\mathcal{L}_{Z\bar e_i e_j}=\bar{e}_i\Big\{\frac{i}{2}\Big[\Big(-g_1 \cos\theta'_W\sin\theta_W+g_2\cos\theta_W\cos\theta'_W+g_{YX}\sin\theta'_W\Big)\sum_{a=1}^3 U^{e,*}_{L,ja}U^{e}_{L,ia}\nonumber\\
&&\hspace{1.6cm}+2\Big((-g_X Q_a+g_{YX})\sin\theta'_W-g_1\cos\theta'_W\sin\theta_W\Big) U^{e,*}_{L,j4}U^{e}_{L,i4}\Big]\gamma_\mu P_L\nonumber\\
&&\hspace{1.6cm}+i\Big[\Big(g_{YX}\sin\theta'_W-g_1\cos\theta'_W\sin\theta_W\Big)\sum_{a=1}^3 U^{e,*}_{R,ia}U^{e}_{R,ja}\nonumber\\
&&\hspace{1.6cm}+\Big((g_XQ_b+g_{YX})\sin\theta'_W-g_1\cos\theta'_W\sin\theta_W\Big) U^{e,*}_{R,i4}U^{e}_{R,j4}\Big]\gamma_\mu P_R\Big\}e_j Z_\mu,
\end{eqnarray}
\begin{eqnarray}
&&\mathcal{L}_{Z\bar d_i d_j}=\bar{d}_i\Big\{\frac{i}{6}\Big[\Big(3g_2 \cos\theta_W\cos\theta'_W+g_1\cos\theta'_W\sin\theta_W-g_{YX}\sin\theta'_W\Big)\sum_{a=1}^3 U^{d,*}_{L,ja}U^{d}_{L,ia}\nonumber\\
&&\hspace{1.6cm}+2\Big((-3g_X Q_a+g_{YX})\sin\theta'_W-g_1\cos\theta'_W\sin\theta_W\Big) U^{d,*}_{L,j4}U^{d}_{L,i4}\Big]\gamma_\mu P_L\nonumber\\
&&\hspace{1.6cm}+\frac{i}{3}\Big[\Big(g_{YX}\sin\theta'_W-g_1\cos\theta'_W\sin\theta_W\Big)\sum_{a=1}^3 U^{d,*}_{R,ia}U^{d}_{R,ja}\nonumber\\
&&\hspace{1.6cm}+\Big((3g_XQ_b+g_{YX})\sin\theta'_W-g_1\cos\theta'_W\sin\theta_W\Big) U^{d,*}_{R,i4}U^{d}_{R,j4}\Big]\gamma_\mu P_R\Big\}d_j Z_\mu.
\end{eqnarray}

To save space in the text, the remaining vertices employed in our calculation are compiled in the Appendix~\ref{A1}.
\section{the processes $h\rightarrow Z\gamma$ and $h\rightarrow m_V Z$}\label{sec3}
\begin{figure}[ht]
\setlength{\unitlength}{5.0mm}
\centering
\includegraphics[width=6.5in]{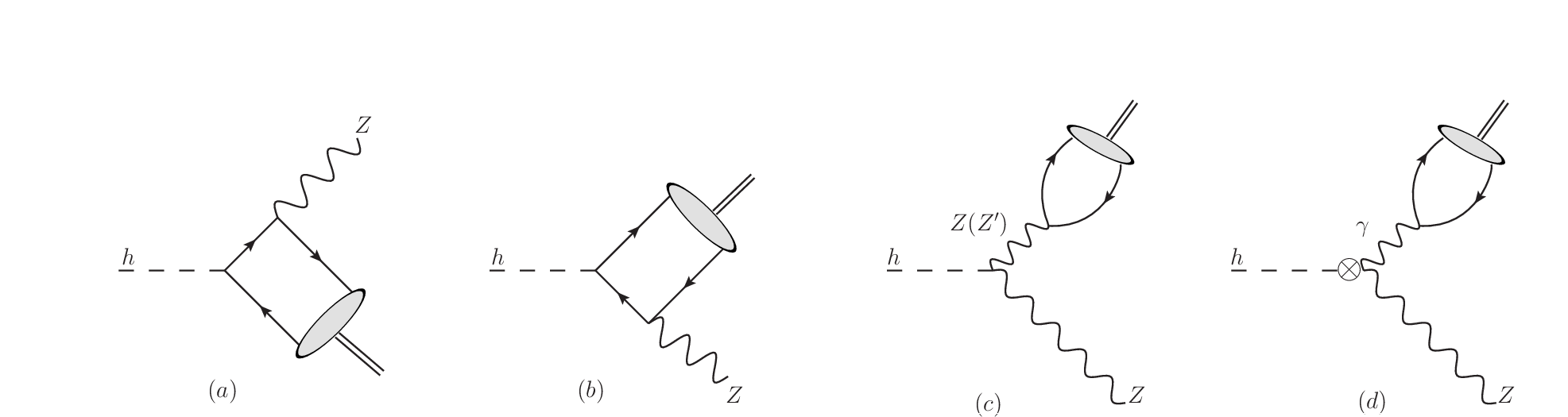}
\caption{The diagrams contributing to the decay $h\rightarrow m_V Z$.}\label{N1}
\end{figure}

This section provides the analytical expressions for the decay $h\to Z\gamma$ and the weak hadronic Higgs decay $h\to m_V Z$. The representative Feynman diagrams of the process $h\to m_V Z$ are shown in Fig.~\ref{N1}. Figs.~\ref{N1}(a) and \ref{N1}(b) correspond to the direct contributions, while Figs.~\ref{N1}(c) and \ref{N1}(d) denote the indirect ones. In loop induced topologies, the effective vertex $h\to Z\gamma^*$ is represented by a crossed circle. As discussed in Ref.~\cite{htomz}, the direct contributions originate from the coupling between the Higgs boson and the constituent quarks of the final state vector meson. Although these diagrams appear at tree level, they usually give only a subdominant contribution. The indirect contributions proceed through the decay process $h\to ZZ^*/Z\gamma^* \to m_V Z$, where the off-shell bosons $Z^*$ or $\gamma^*$ hadronize into the vector meson. In the topology of Fig.~\ref{N1}(c), the virtual $Z$ can in principle be replaced by a $Z'$. However, current limits require $m_{Z'}>5.15~\text{TeV}$, making its effect negligible. Numerically, $|\frac{1}{m^2_V-m^2_Z}| \sim \frac{1}{90^2~\rm GeV^2}$ and $|\frac{1}{m^2_V-m^2_{Z'}}| \sim \frac{1}{5100^2~\rm GeV^2}$, indicating that the latter is approximately $10^{-4}$ times the former; therefore the $Z'$ exchange is omitted in our analysis. Among the indirect contributions, the decay $h\to ZZ^*$ occurs at tree level in the SM, while the $h\gamma Z$ interaction is loop induced. In the $U(1)_X$VLFM framework, this vertex receives additional nonstandard contributions. The corresponding effective Lagrangian is given by
\begin{eqnarray}
\mathcal{L}_{eff}= \frac{\alpha}{4 \pi v} \Big(\frac{2 C_{\gamma Z}}{\sin\theta_{W} \cos\theta_{W}} h F_{\mu v} Z^{\mu v}-\frac{2 {\tilde{C}}_{\gamma Z}}{\sin\theta_{W} \cos\theta_{W}} h F_{\mu v} {\tilde{Z}}^{\mu v}\Big).\label{f1}
\end{eqnarray}
Here, $\theta_W$ is the weak mixing angle. Using the effective Lagrangian in Eq.(\ref{f1}), the decay width of $h \rightarrow Z\gamma$ is derived
\begin{eqnarray}
\Gamma(h \rightarrow Z \gamma)= \frac{{\alpha}^2 m^3_h}{32 {\pi}^3 v^2 \sin^2\theta_{W} \cos^2\theta _{W}}\Big(1-\frac{m^2_Z}{m^2_h}\Big)^3 (|C_{\gamma Z}|^2+|{\tilde{C}}_{\gamma Z}|^2).
\end{eqnarray}

Loop diagrams associated with NP can induce additional corrections to the processes $h\to m_V\gamma$ and $h\to m_V Z$. Although these decays share similar topologies, their physical properties differ significantly. In the case of $h\to m_V\gamma$, the photon in the final state is on shell and massless, and therefore does not possess a longitudinal polarization component. Within the NRQCD framework, the gauge invariant amplitude for $h\to m_V\gamma$ at tree level can be written as
\begin{eqnarray}
&&{\cal M}^{\gamma} = \frac{4\sqrt{3} e e_q\phi_0}{m^2_{h} - m^2_V}\Big(\frac{m_VG_F}{2\sqrt{2}}\Big)^{\frac{1}{2}}
\Big[c_S\{2(\varepsilon^*_\gamma\cdot p_V)(\varepsilon^*_V\cdot k_\gamma) - (m^2_{h} - m^2_V)(\varepsilon^*
_\gamma\cdot\varepsilon^*_V)\}\nonumber\\
&&-2c_P\epsilon_{\mu\nu\rho\lambda}~\varepsilon^{*\mu}_\gamma k_\gamma^\nu p^\rho
_V\varepsilon^{*\lambda}_V\Big]\label{gs1}.
\end{eqnarray}
Here, $k_\gamma (p_V)$ denotes the four-momentum of the photon (vector meson), and $\varepsilon_\gamma^* (\varepsilon_V^*)$ is the corresponding polarization vector.

In the rest frame of the vector meson, Eq.~(\ref{gs1}) can be recast into a more transparent form. Using the definitions
$\varepsilon^{*L}_V\equiv \bar{\varepsilon}^{*}_V\cdot\hat{k}_\gamma$ and $\bar{\varepsilon}^{*T}_V\equiv \bar{\varepsilon}^{*}_V-\varepsilon^{*L}_V\hat{k}_\gamma$,
the decay amplitude in the transverse basis becomes \cite{HQQ1}
\begin{eqnarray}
&&{\cal M}^\gamma = H^{\gamma}_{\parallel}{\vec\varepsilon}^{*T}_V\cdot\vec\varepsilon^*_\gamma + iH^{\gamma}_\perp
{\hat k}_\gamma\cdot({\vec\varepsilon}^{*T}_V\times\vec\varepsilon^*_\gamma), \label{gs2}\\&&
H^{\gamma}_{\parallel}= 4\sqrt{3} e e_q\phi_0\Big(\frac{m_VG_F}{2\sqrt{2}}\Big)^{\frac{1}{2}} c_S,\\
&& H^{\gamma}_\perp= 4\sqrt{3} e e_q\phi_0\Big(\frac{m_VG_F}{2\sqrt{2}}\Big)^{\frac{1}{2}} ic_P.
\label{Hperpdef}
\end{eqnarray}
Because the photon in the final state is both massless and on shell, longitudinal polarization does not contribute to the amplitude and only transverse polarizations appear in Eq.~(\ref{gs2}). The triple product
${\hat k}_\gamma\cdot({\vec\varepsilon}^{*T}_V\times\vec\varepsilon^*_\gamma)$
constitutes the unique P-odd observable in $|\mathcal{M}|^2$, whose coefficient is proportional to $c_Sc_P$. The parameter $c_P$ encodes a pseudoscalar $H q\bar q$ coupling arising from possible NP effects and manifests itself through a nonzero value of this triple product. However, since the photon does not decay, its polarization cannot be reconstructed experimentally, making a direct measurement of $c_P$ impossible.

To overcome the limitation imposed by the unobservable photon polarization, one may instead consider replacing the photon with a massive vector boson $Z$, whose polarization can be measured through its decay. Unlike the photon, the $Z$ boson couples to the Higgs already at tree level, introducing an additional diagram that contributes to the decay $h\to m_V Z$. Although the $Z\bar q q$ interaction contains an axial-vector component, this term does not contribute to the leading NRQCD matrix element for $h\to m_V Z$. In the rest frame of the vector meson, we choose $\hat{k}_Z$ as the direction of the outgoing $Z$ boson. Following an analysis analogous to the $h\to m_V\gamma$ case, the decay amplitude can be written in the transverse-longitudinal basis as
\begin{eqnarray}
&&{\cal M}^Z = H^Z_{0}{\vec\varepsilon}^{*L}_V\cdot\vec\varepsilon^{*L}_Z+H^Z_{\parallel}{\vec\varepsilon}^{*T}_V\cdot\vec\varepsilon^{*T}_Z + iH^Z_\perp
{\hat k}_Z\cdot({\vec\varepsilon}^{*T}_V\times\vec\varepsilon^{*T}_Z).
\end{eqnarray}
The coefficients $H^Z_0$ and $H^Z_{\parallel}$ scale with the scalar coupling $c_S$, while $H^Z_\perp$ is proportional to the pseudoscalar parameter $c_P$. Explicit expressions for these form factors can be found in Ref.~\cite{HQQ1}. Since the polarization of the $Z$ boson can be inferred from its decay products, the transverse component $\varepsilon^{*T}_Z$ is experimentally accessible. A nonvanishing triple-product term would therefore constitute clear evidence for a pseudoscalar contribution and provide a direct probe of $c_P$.

For the decay $h\rightarrow m_V Z$, the amplitude is commonly expressed in terms of longitudinal and transverse polarization components. A convenient parametrization reads
\begin{equation}
   i{\cal A}(h\to m_VZ)
   = - \frac{2g m_V}{\cos\theta_W \upsilon}
    \left[ \varepsilon_V^{\parallel *}\cdot\varepsilon_Z^{\parallel *}\,F_\parallel^{VZ}
    + \varepsilon_V^{\perp *}\cdot\varepsilon_Z^{\perp *}\,F_\perp^{VZ}
    + \frac{\epsilon_{\mu\nu\alpha\beta}\,k_V^\mu k_Z^\nu\varepsilon_V^{*\alpha}\varepsilon_Z^{*\beta}}
           {\left[ (k_V\cdot k_Z)^2-k_V^2 k_Z^2\right]^{1/2}}\,\widetilde F_\perp^{VZ} \right] ,
\end{equation}
where $k_Z$ is the four-momentum of the outgoing $Z$ boson. The longitudinal and transverse polarization vectors of the vector meson are defined by \cite{htomz}
\begin{equation}
   \varepsilon_V^{\parallel \mu} = \frac{1}{m_V}\,\frac{k_V\cdot k_Z}{\left[ (k_V\cdot k_Z)^2-k_V^2 k_Z^2\right]^{1/2}}
    \left( k_V^\mu - \frac{k_V^2}{k_V\cdot k_Z}\,k_Z^\mu \right) , \qquad
   \varepsilon_V^{\perp\mu} = \varepsilon_V^\mu - \varepsilon_V^{\parallel \mu}.\label{vectorV}
\end{equation}
The polarization vectors of the $Z$ boson follow from Eq.(\ref{vectorV}) by performing the replacements $m_V\rightarrow m_Z$ and $k_V\leftrightarrow k_Z$.

For the decay $h\rightarrow m_V Z$, the partial width can be written as
\begin{eqnarray}
&&\hspace{-0.5cm}\Gamma(h\rightarrow m_V Z)= \frac{m^3_h}{4 \pi v^4} {\lambda}^{1/2}(1,r_Z,r_V)(1-r_Z-r_V)^2 \nonumber\\&&\hspace{2cm}{\times} \Big[|F^{VZ}_{\parallel}|^2+{\frac{8 r_Z r_V}{(1-r_Z-r_V)^2}} (|F^{VZ}_{\perp}|^2+|\tilde{F}^{VZ}_{\perp}|^2)\Big],\label{f4}
\end{eqnarray}
where $\lambda(x,y,z)=(x-y-z)^2-4 y z$, $r_Z=m^2_Z/m^2_{h}$ and $r_V=m^2_V/m^2_{h}$. Although $r_V\ll 1$ for light vector mesons, the transverse amplitudes exhibit a notable sensitivity to this small parameter. To avoid losing such effects, we retain $r_V=m^2_V/m^2_{h}$ rather than employing a massless approximation, which leads to a more accurate description of the transverse polarization contributions.

In Eq.(\ref{f4}), the three form factors $F^{VZ}_{\parallel}$, $F^{VZ}_{\perp}$ and $\tilde{F}^{VZ}_{\perp}$ each receive contributions from both direct and indirect mechanisms.
For convenience, we first present the indirect parts, which arise from the effective $h\gamma Z$ vertex and the tree-level $hZZ$ coupling, and their expressions are given as follows
\begin{eqnarray}
  &&   F_{\parallel\,\rm indirect}^{VZ} = \frac{\kappa_Z}{1-r_V/r_Z} \sum_q f_V^q\,v_q
    + C_{\gamma Z}\,\frac{\alpha(m_V)}{4\pi}\,\frac{4r_Z}{1-r_Z-r_V} \sum_q f_V^q\,Q_q,  \nonumber\\&&
   F_{\perp\,\rm indirect}^{VZ} = \frac{\kappa_Z}{1-r_V/r_Z} \sum_q f_V^q\,v_q
    + C_{\gamma Z}\,\frac{\alpha(m_V)}{4\pi}\,\frac{1-r_Z-r_V}{r_V} \sum_q f_V^q\,Q_q, \nonumber\\&&
   \widetilde F_{\perp\,\rm indirect}^{VZ}
   = \widetilde C_{\gamma Z}\,\frac{\alpha(m_V)}{4\pi}\,
    \frac{\lambda^{1/2}(1,r_Z,r_V)}{r_V} \sum_q f_V^q\,Q_q.\label{FVZindirect}
\end{eqnarray}
The vector and axial-vector couplings of $Z \bar{q} q$ are expressed as $v_q =\frac{T^q_3}{2}-Q_q \sin^2\theta_{W}$ and $a_q =\frac{T^q_3}{2}$, respectively. The vector meson decay constant $f^q_V$ is introduced through
\begin{eqnarray}
\Big<V(k,\varepsilon)|\bar{q} {\gamma}^{\mu} q| 0\Big>= -i f^q_V m_V {\varepsilon}^{* \mu},~~~q = u, d, s\dots
\end{eqnarray}
To calculate the results, we make use of the relations
\begin{eqnarray}
Q_V f_V=\sum_{q} Q_q f^q_V,~~~~\sum_{q} f^q_V v_q=f_V v_V.
\end{eqnarray}

The specific forms of $C_{\gamma Z}$ and $\tilde{C}_{\gamma Z}$ in Eq.(\ref{FVZindirect}) are as follows~\cite{CGZ}
\begin{eqnarray}
&&C_{\gamma Z}=C^{SM,light}_{\gamma Z}+C^{U(1)_X}_{\gamma Z},~~~~\tilde{C}_{\gamma Z}=\tilde{C}^{SM,light}_{\gamma Z}+\tilde{C}^{U(1)_X}_{\gamma Z},
\nonumber\\&&C^{SM,light}_{\gamma Z}= \sum_{q=u,d,c,s} \frac{2 N_c Q_q v_q}{3} A_f({\tau}_q,r_Z)+\sum_{l=\mu,e} \frac{2 Q_l v_l}{3} A_f({\tau}_l,r_Z)- \frac{1}{2} A^{\gamma Z}_W({\tau}_W,r_Z),
\nonumber\\&&\tilde{C}^{SM,light}_{\gamma Z}= \sum_{q=u,d,c,s} \tilde{\kappa}_q N_c Q_q v_q B_f({\tau}_q,r_Z)+\sum_{l=\mu,e} \tilde{\kappa}_l Q_l v_l B_f({\tau}_l,r_Z),
\end{eqnarray}
with ${\tau}_i=4 m^2_i/m^2_{h}$. $C^{SM,light}_{\gamma Z}$ and $\tilde{C}^{SM,light}_{\gamma Z}$ correspond to the first two generation SM fermions and $W$ gauge boson contributions to $h\rightarrow Z \gamma$, while $A_f$,~$B_f$ and $A^{\gamma Z}_W$ are all loop functions~\cite{htomgamma1,htomgamma2,htomgamma3}.

\begin{figure}[ht]
\setlength{\unitlength}{5.0mm}
\centering
\includegraphics[width=2.0in]{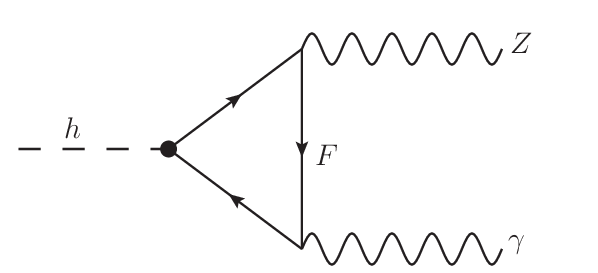}
\caption{The one loop diagrams with new particles for the decay $h\rightarrow Z \gamma$.}\label{N2}
\end{figure}

Fig.~\ref {N2} shows the NP one loop corrections to the decay $h\rightarrow Z\gamma$ in the $U(1)_X$VLFM model. The new contributions to $C_{\gamma Z}$ arise mainly from the vectorlike leptons and vectorlike quarks introduced in the model. The mixing between the vectorlike fermions and the third generation SM fermions induces additional effects in the fermionic
loop amplitudes, rendering them different from the purely SM contributions. As noted in Ref.~\cite{tt5}, the QCD corrections to the $h\rightarrow Z\gamma$ decay width are at the level of about 0.1$\%$, which is extremely small and can be safely neglected in our analysis.

The extended gauge structure modifies the effective couplings that enter the fermionic loop amplitudes. To illustrate this explicitly, we take the $Z-\bar{u}_i-u_i$ vertex as an example. The corresponding interactions in the SM and $U(1)_X$VLFM are given by
\begin{eqnarray}
&&\mathcal{L}^{SM}_{Z\bar{u}u}=\bar{u}_i\Big\{-\frac{ e}{2 \sin\theta_W \cos\theta_W} \Big(1-\frac{4}{3}{\sin^2\theta_W}\Big){\gamma}_{\mu}P_L+\frac{2e \sin\theta_W}{3\cos\theta_W}{\gamma}_{\mu}P_R \Big\}u_iZ_\mu,\label{s1}
\\&&\mathcal{L}^{U(1)_X}_{Z\bar{u}u}=\bar{u}_i\Big\{-\frac{i}{6}\Big[\Big(3g_2 \cos\theta_W\cos\theta'_W-g_1\cos\theta'_W\sin\theta_W+g_{YX}\sin\theta'_W\Big)\sum_{a=1}^3 U^{u,*}_{L,ja}U^{u}_{L,ia}\nonumber\\
&&\hspace{1.6cm}+2\Big((2g_{YX}-3g_X Q_a)\sin\theta'_W-2g_1\cos\theta'_W\sin\theta_W\Big) U^{u,*}_{L,j4}U^{u}_{L,i4}\Big]\gamma_\mu P_L\nonumber\\
&&\hspace{1.6cm}+\frac{i}{3}\Big[2\Big(g_1\cos\theta'_W\sin\theta_W-g_{YX}\sin\theta'_W\Big)\sum_{a=1}^3 U^{u,*}_{R,ia}U^{u}_{R,ja}\nonumber\\
&&\hspace{1.6cm}+\Big(2g_1\cos\theta'_W\sin\theta_W-(3g_XQ_b+2g_{YX})\sin\theta'_W\Big) U^{u,*}_{R,i4}U^{u}_{R,j4}\Big]\gamma_\mu P_R\Big\}u_j Z_\mu,\label{ss}
\end{eqnarray}
where the couplings in Eq.(\ref{ss}) depend on $\theta_W'$, $g_X$ and $g_{YX}$, with $\theta_W'$ originating from the $Z$-$Z'$ mixing.

Setting $\theta_W'=0$ and assuming no mixing between the three SM up-type quarks and the vectorlike fourth generation, Eq.(\ref{ss}) reduces to
\begin{eqnarray}
&&\mathcal{L}^{U(1)_X}_{Z\bar{u}u}\rightarrow\frac{1}{6}\bar{u}_i\Big[(-3 g_2\cos\theta_W+g_1\sin\theta_W){\gamma}_{\mu}P_L+4g_1\sin\theta_W{\gamma}_{\mu}P_R \Big]u_i Z^{\mu}
\nonumber\\&&\hspace{1.3cm}=\bar{u}_i\Big[-\frac{ e}{2 \sin\theta_W \cos\theta_W} (1-\frac{4}{3}{\sin^2\theta_W}){\gamma}_{\mu}P_L+\frac{2e \sin\theta_W}{3\cos\theta_W}{\gamma}_{\mu}P_R \Big]u_i Z^{\mu}.\label{kk}
\end{eqnarray}
which is identical to the SM result in Eq.(\ref{s1}).

Our numerical analysis shows that the typical value of the mixing angle is $\theta_W' \sim 10^{-5}$ in the $U(1)_X$VLFM, leading to a relative deviation $\frac{\mathcal{L}^{SM}_{Z\bar{u}u}-\mathcal{L}^{U(1)_X}_{Z\bar{u}u}}{\mathcal{L}^{SM}_{Z\bar{u}u}}$ is at the order of $10^{-5}$. Such a tiny difference has a negligible impact on the loop induced amplitudes. Therefore, for fermionic loops involving SM fields, we safely adopt the SM couplings in our calculations.

In the $U(1)_X$VLFM, the CP-even coupling $C^{U(1)_X}_{\gamma Z}$ reads
\begin{eqnarray}
&&\hspace{-0.3cm}C^{U(1)_X}_{\gamma Z}=\frac{v \sin\theta_W \cos\theta_W}{e}\int_{0}^{1}dx\int_{0}^{1}ydy  \sum_{F=t,b,\tau,t',b',\tau'}\Big[ \frac{Q_{F1}}{R_{1N}^2(m_{F_1},m_{F_2})}\Big(A^{\bar{F}_2 F_1 h} B^{\bar{F}_1 F_2 Z}\nonumber\\&&\hspace{1.0cm}\times(-2(x-1)y^2(m_{F_1}+m_{F_2})+y(2 x(m_{F_1}+m_{F_2})-3 m_{F_1}-m_{F_2})+m_{F_1})\nonumber\\&&\hspace{1.0cm}+A_w^{\bar{F}_2 F_1 h} B_w^{\bar{F}_1 F_2 Z}(m_{F_1}(y-1)(2(x-1)y+1)+ym_{F_2}(-2xy+2x+2y-1))\Big)\nonumber\\&&\hspace{1.0cm}+\frac{Q_{F1}}{R_{2N}^2(m_{F_1},m_{F_2})}\Big(A^{\bar{F}_2 F_1 h} B^{\bar{F}_1 F_2 Z}(-2(x-1)y^2(m_{F_1}+m_{F_2})+y(x-1)\nonumber\\&&\hspace{1.0cm}\times(3m_{F_1}+m_{F_2})+m_{F_1})+A_w^{\bar{F}_2 F_1 h} B_w^{\bar{F}_1 F_2 Z}(m_{F_1}(y-1)(2(x-1)y+1)\nonumber\\&&\hspace{1.0cm}-y(m_{F_1}x+m_{F_2}(x-1)(2y-1)))\Big)\Big].
\end{eqnarray}

The explicit expression of $C^{U(1)_X}_{\gamma Z}$ is given by
\begin{eqnarray}
&&\hspace{-0.3cm}\tilde{C}^{U(1)_X}_{\gamma Z}=-\frac{i v \sin\theta_W \cos\theta_W}{e}\int_{0}^{1}dx \int_{0}^{1}ydy \sum_{F=t,b,\tau,t',b',\tau'}\Big[\frac{1}{R_{1N}^2(m_{F_1},m_{F_2})} \nonumber\\&&\hspace{1.0cm}\times\Big(A^{\bar{F}_2 F_1 h} B_w^{\bar{F}_1 F_2 Z}(y(m_{F_1}+m_{F_2})-m_{F_1})+A_w^{\bar{F}_2 F_1 h} B^{\bar{F}_1 F_2 Z}(m_{F_1}(1-y)+m_{F_2}y)\Big)\nonumber\\&&\hspace{1.0cm}+\frac{1}{R_{2N}^2(m_{F_1},m_{F_2})} \Big(A^{\bar{F}_2 F_1 h} B_w^{\bar{F}_1 F_2 Z}(y(1-x)(m_{F_1}+m_{F_2})-m_{F_1})+A_w^{\bar{F}_2 F_1 h} B^{\bar{F}_1 F_2 Z}\nonumber\\&&\hspace{1.0cm}\times((x-1)y(m_{F_1}-m_{F_2})+m_{F_1})\Big)\Big].
\end{eqnarray}

The functions $R_{1N}^2(m_{1},m_{2})$ and $R_{2N}^2(m_{1},m_{2})$ are shown as
\begin{eqnarray}
&&R_{1N}^2(m_{1},m_{2})=p^2_2(1-x)^2 y^2+p^2_1(1-y)^2-2 p_1{\cdot}p_2(1-x)y(1-y)+m^2_2xy\nonumber\\&&\hspace{2.8cm}+(m^2_2-p^2_2)(1-x)y+(m^2_1-p^2_1)(1-y),
\nonumber\\&&R_{2N}^2(m_{1},m_{2})=p^2_1(1-x)^2 y^2+p^2_2(1-y)^2-2 p_1{\cdot}p_2(1-x)y(1-y)+m^2_2xy\nonumber\\&&\hspace{2.8cm}+(m^2_2-p^2_1)(1-x)y+(m^2_1-p^2_2)(1-y).
\end{eqnarray}

The interaction between $F_1$, $F_2$, and the Higgs boson is parametrized by the scalar and pseudoscalar couplings $A^{\bar F_2 F_1 h}$ and $A_w^{\bar F_2 F_1 h}$.
The $Z$-boson couplings to $\bar F_1 F_2$ are encoded in the vector and axial-vector coefficients $B^{\bar F_1 F_2 Z}$ and $B_w^{\bar F_1 F_2 Z}$. These interactions can be written in the generic form
\begin{eqnarray}
\bar{F}_2 i (A^{\bar{F}_2 F_1 h}+A_w^{\bar{F}_2 F_1 h} {\gamma}_{5}) F_1 h,~~~
 \bar{F}_1 i(B^{\bar{F}_1 F_2 Z} {\gamma}_{\mu}+B_w^{\bar{F}_1 F_2 Z} {\gamma}_{\mu} {\gamma}_{5})F_2 Z^{\mu}.
\end{eqnarray}
All explicit expressions for the relevant vertices are provided in Sec.~\ref{sec2} and Appendix~\ref{A1}.

The direct and indirect contributions behave very differently: the former can only be evaluated as an expansion in power series of $(m_q/m_{h})^2$ or $(\Lambda_{QCD}/m_{h})^2$, where $m_q$ denotes the constituent quark mass inside the meson and $\Lambda_{QCD}$ characterizes the hadronic scale.
For a longitudinally polarized vector meson, the direct contribution originates from subleading-twist components and is therefore power suppressed. In contrast, for a transversely polarized vector meson, the leading-twist distribution amplitude enters directly. Using the asymptotic form $\phi_V^\perp(x)=6x(1-x)$~\cite{GZBHS1,GZBHS2,GZBHS3}, the direct contributions take the form
\begin{eqnarray}
   F_{\perp\,{\rm direct}}^{VZ}
   &= \sum_q f_V^{q\perp} v_q\,\kappa_q\,\frac{3m_q}{2m_V}\,\frac{1-r_Z^2+2r_Z\ln r_Z}{(1-r_Z)^2} \,, \\
   \widetilde F_{\perp,{\rm direct}}^{VZ}
   &= \sum_q f_V^{q\perp} v_q\,\tilde\kappa_q\,\frac{3m_q}{2m_V}\,\frac{1-r_Z^2+2r_Z\ln r_Z}{(1-r_Z)^2} \,.
\end{eqnarray}
Although these expressions may appear numerically comparable to the indirect term in Eq.(\ref{FVZindirect}), the direct contributions remain strongly suppressed once the small quark masses are taken into account.
\section{numerical analysis}\label{sec4}
In this section, we impose several experimental constraints on the parameter space of the $U(1)_X$VLFM:

1. The mass of the lightest CP-even Higgs boson is fixed to $m_h = 125.20 \pm 0.11~{\rm GeV}$ \cite{PDG}.

2. After mixing between the third-generation SM fermions and the vectorlike states, the physical masses are required to reproduce the SM values: $m_t = 172.57 \pm 0.29~{\rm GeV}, m_b = 4.183 \pm 0.007~{\rm GeV}, m_\tau = 1.78 \pm 0.09~{\rm GeV}$ \cite{PDG}.

3. The latest ATLAS and CMS results set 95$\%$ CL lower mass bounds of about 1.49-1.52 TeV for vectorlike quarks~\cite{Benbrik:2024fku}. For vectorlike leptons, CMS excludes long-lived scenarios below 700 GeV~\cite{CMS:2025urb}, while ATLAS electroweak searches exclude masses below 910 GeV~\cite{ATLAS:2025wgc}.

4. The additional gauge boson satisfies $M_{Z'}\geq 5.15~ {\rm TeV}$ at 95$\%$ CL~\cite{Zpupper}.
Moreover, the ratio $\frac{M_{Z'}}{g_X}$ is constrained to be larger than 6~{\rm TeV} at 99$\%$ CL~\cite{Zpupper1,Zpupper2}, which limits the gauge coupling to $0< g_X \leq0.85$.

After applying these experimental requirements, a sizable set of viable sample points is obtained, from which one-dimensional distributions and multidimensional scatter plots are constructed.

The numerical analysis is organized into four parts: (1) determination of the relevant input parameters; (2) analysis of the decays $h\to\gamma\gamma$ and $h\to VV^*(V=Z,W)$; (3) discussion of the $h\to Z\gamma$ decay; (4) study of the processes $h\to m_V Z$, where $m_V=\omega,\rho,\phi,J/\psi,\Upsilon$.
\subsection{ The input parameters scheme }
Under the constraints from quark and charged-lepton masses, the Yukawa couplings for the first two generations are taken as
\begin{eqnarray}
&&Y_{u_i}=\sqrt{2}\, m_{u_i}/v,\qquad
Y_{d_i}=\sqrt{2}\, m_{d_i}/v,\qquad
Y_{e_i}=\sqrt{2}\, m_{l_i}/v\quad (i=1,2),
\end{eqnarray}
while the third-generation Yukawa couplings are determined by the mixing with the vectorlike fermions, given by
\begin{eqnarray}
&&Y_{u_3} = \frac{\sqrt{2}\, m_t\sqrt{\,2m_t^2-v_P^2 Y_{PU}^2 - v_S^2 Y_{XU}^2\,}}
{v\sqrt{\,2m_t^2-v_P^2 Y_{PU}^2\,}},\nonumber\\&&
Y_{d_3} = \frac{\sqrt{2}\, m_b\sqrt{\,2m_b^2-v_P^2 Y_{PD}^2 - v_S^2 Y_{XD}^2\,}}
{v\sqrt{\,2m_b^2-v_P^2 Y_{PD}^2\,}},\nonumber\\&&
Y_{e_3} = \frac{\sqrt{2}\, m_\tau\sqrt{\,2m_\tau^2-v_P^2 Y_{PE}^2 - v_S^2 Y_{XE}^2\,}}
{v\sqrt{\,2m_\tau^2-v_P^2 Y_{PE}^2\,}}.
\end{eqnarray}
Here $m_{u_i}$, $m_{d_i}$ and $m_{l_i}$ denote the up-type quark, down-type quark and charged-lepton masses, respectively, and their values are taken from the latest PDG~\cite{PDG}.

The following model parameters are fixed throughout the analysis
\begin{eqnarray}
&&Q_a = 1,\quad Q_b = 1,\quad L_H=-0.14,\quad L_P=-0.1,\quad L_X=-0.06,\quad\nonumber\\&&
L_{HP}=-0.01,\quad L_{HX}=-0.05,\quad L_{PX}=-0.05.
\end{eqnarray}

For the numerical study, we scan over the parameter set
\begin{eqnarray}
&&g_X,\quad g_{YX},\quad Y_{XD},\quad Y_{PD},\quad Y_{XU},\nonumber\\&&
Y_{PU},\quad Y_{XE},\quad Y_{PE},\quad v_P,\quad v_S,
\end{eqnarray}
which contains the dominant inputs affecting the predictions of $h\to Z\gamma$ and $h\to m_VZ$ in the $U(1)_X$VLFM.

\begin{table}[ht]
\caption{Input values for the vector meson decay constants}
\begin{tabular}{|c|c|c|c|c|c|}
\hline
Vector meson & $\omega$ & $\rho$ & $\phi$ & $J/\psi$ & $\Upsilon$ \\
\hline
$m_V/{\rm GeV}$ & 0.782 & 0.77 & 1.02 & 3.097 & 9.46 \\
\hline
$f_V/{\rm GeV}$ & 0.194 & 0.216 & 0.223 & 0.403 & 0.684 \\
\hline
$\upsilon_V$ &$ -\frac{\sin^2\theta_W}{3 \sqrt{2}}$ & $\frac{1}{\sqrt{2}} (\frac{1}{2}-\sin^2\theta_W)$ & $-\frac{1}{4}+\frac{\sin^2\theta_W}{3}$ & $\frac{1}{4}-\frac{2 \sin^2\theta_W}{3}$ & $-\frac{1}{4}+\frac{\sin^2\theta_W}{3}$  \\
\hline
$Q_V$ & $\frac{1}{3 \sqrt{2}}$ & $\frac{1}{\sqrt{2}}$ & $-\frac{1}{3}$ & $\frac{2}{3}$ & $-\frac{1}{3}$ \\
\hline
${f^{\perp}_V}/{f_V}={f^{q \perp}_V}/{f^{q}_V}$ & 0.71 & 0.72 & 0.76 & 0.91 & 1.09 \\
\hline
\end{tabular}
\label{t1}
\end{table}

In addition, the numerical inputs for the vector meson decay parameters, including $m_{V}$, $f_{V}$, $v_{V}$, $Q_V$ and the ratio ${f^{\perp}_V}/{f_V}={f^{q \perp}_V}/{f^{q}_V}$, are compiled in Table \ref{t1}. Here, ${f^{\perp}_V}$ and ${f^{q \perp}_V}$ denote the transverse decay constants and the flavor-specific transverse decay constants, respectively.

\subsection{ The processes $h\rightarrow \gamma \gamma$ and $h\rightarrow V V^*$ }
In this section, under the parameter choice $g_X = 0.6,\; g_{YX} = -0.1,\; Y_{XE} = 0.5,\; Y_{PE} = 0.005,\; v_P = 4500~\text{GeV},\; v_S = 8500~\text{GeV}$, we evaluate the ratios $R_{\gamma\gamma}$ and $R_{VV^*} (V = Z, W)$ corresponding to the decay processes $h\to\gamma\gamma$ and $h\to V V^*$, respectively.

\begin{figure}[ht]
\setlength{\unitlength}{5mm}
\centering
\includegraphics[width=2.9in]{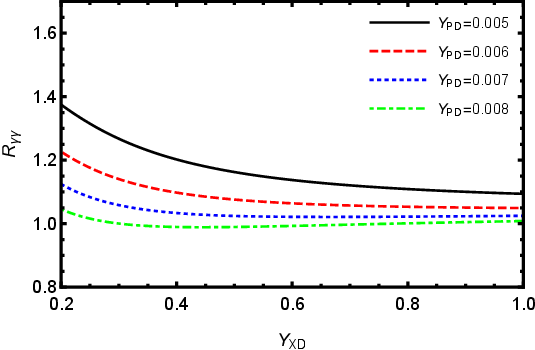}
\caption{$R_{\gamma\gamma}$ varying with $Y_{XD}$ for different $Y_{PD}$.}{\label {r2gamma}}
\end{figure}

Setting $Y_{XU}=0.5$ and $Y_{PU}=0.07$, Fig.~\ref{r2gamma} illustrates how $R_{\gamma\gamma}$ varies as a function of $Y_{XD}$. The black, red, blue and green curves correspond to $Y_{PD}=0.005,\ Y_{PD}=0.006,\ Y_{PD}=0.007$ and $Y_{PD}=0.008$, respectively. As observed in the Fig.~\ref{r2gamma}, all four curves exhibit a mildly decreasing trend as $Y_{XD}$ increases, and they gradually level off at larger values of $Y_{XD}$. Moreover, increasing $Y_{PD}$ systematically lowers the overall height of the curves. In the small $Y_{XD}$ region, especially for smaller $Y_{PD}$ the predicted values of $R_{\gamma\gamma}$ are generally above the experimental central value. As $Y_{XD}$ increases, however, all four curves gradually decrease and converge toward the vicinity of 1.1, leading to improved agreement between the theoretical predictions and the experimental measurement.

\begin{figure}[ht]
\setlength{\unitlength}{5mm}
\centering
\includegraphics[width=2.9in]{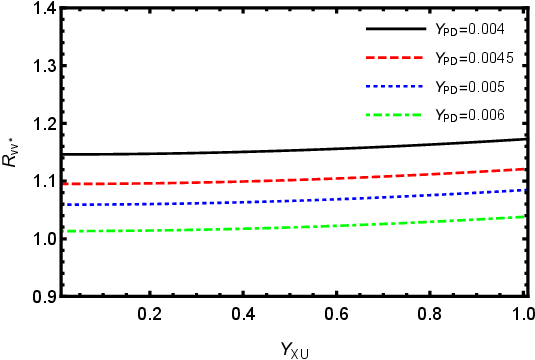}
\caption{$R_{VV^*}$ varying with $Y_{XU}$ for different $Y_{PD}$.}{\label {r2v}}
\end{figure}

Based on the inputs $Y_{XD}=0.5$ and $Y_{PU}=0.055$, the black ($Y_{PD}=0.004$), red ($Y_{PD}=0.0045$), blue ($Y_{PD}=0.005$) and green ($Y_{PD}=0.006$) curves in Fig.~\ref{r2v} present the dependence of $R_{VV^*}$ on $Y_{XU}$. Overall, all four curves exhibit a mild upward trend: as $Y_{XU}$ increases, $R_{VV^*}$ rises gradually and monotonically, with a very small slope. This indicates that the impact of $Y_{XU}$ on this ratio is positive but relatively weak. In addition, smaller values of $Y_{PD}$ shift the entire curve upward. The four curves are nearly parallel and show only minimal fluctuations throughout the plotted range, with their values confined to the narrow interval of approximately 1.03-1.15.

Since the parameter choices satisfy the Higgs experimental constraints, we no longer display the results for $R_{\gamma\gamma}$ and $R_{VV^*} (V = Z, W)$ in the subsequent numerical analysis.
\subsection{ The process $h\rightarrow Z \gamma$ }
\begin{figure}[ht]
\setlength{\unitlength}{5mm}
\centering
\includegraphics[width=2.9in]{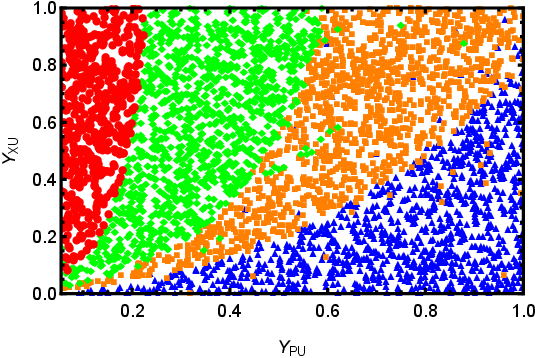}
\caption{${\Gamma}_{NP}(h\rightarrow Z \gamma)/{\Gamma}_{SM}(h\rightarrow Z \gamma)$ in $Y_{PU}-Y_{XU}$ plane, and the marking of the scattering points represents: $\textcolor{blue}{\blacktriangle}~({\Gamma}_{NP}(h\rightarrow Z \gamma)/{\Gamma}_{SM}(h\rightarrow Z \gamma)<1.294),~\textcolor{orange}{\blacksquare}~(1.294\leq {\Gamma}_{NP}(h\rightarrow Z \gamma)/{\Gamma}_{SM}(h\rightarrow Z \gamma)<1.3),~ \textcolor{green}{\blacklozenge}~(1.3\leq {\Gamma}_{NP}(h\rightarrow Z \gamma)/{\Gamma}_{SM}(h\rightarrow Z \gamma)<1.35),~ \textcolor{red}{\bullet}~(1.35\leq {\Gamma}_{NP}(h\rightarrow Z \gamma)/{\Gamma}_{SM}(h\rightarrow Z \gamma$).}{\label {htozgamma1}}
\end{figure}

\begin{figure}[ht]
\setlength{\unitlength}{5mm}
\centering
\includegraphics[width=2.9in]{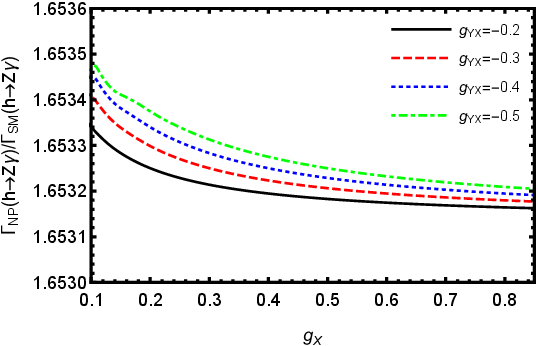}
\caption{${\Gamma}_{NP}(h\rightarrow Z \gamma)/{\Gamma}_{SM}(h\rightarrow Z \gamma)$ varying with $g_X$ for different $g_{YX}$.}{\label {htozgamma2}}
\end{figure}

The NP contribution to the decay $h\rightarrow m_V Z$ originates from the effective $hZ\gamma$ coupling. Therefore, investigating the $h\rightarrow Z\gamma$ process is crucial for probing the properties of the Higgs boson. According to the latest experimental results, the signal strength is measured to be $\mu_{Z\gamma}=2.2\pm0.7$~\cite{hZgexp}.  For the numerical evaluation of the $h\rightarrow Z\gamma$ decay width, we adopt the parameter set $Y_{XD}=0.8,\; Y_{PD}=0.01,\; Y_{XE}=0.5,\; Y_{PE}=0.01,\; v_P=4500~\text{GeV},\; v_S=8500~\text{GeV}$, and perform ${\Gamma}_{NP}(h\rightarrow Z \gamma)/{\Gamma}_{SM}(h\rightarrow Z \gamma)$ schematic diagrams affected by different parameters in the Fig.~\ref{htozgamma1} and Fig.~\ref{htozgamma2}.

Fig.~\ref{htozgamma1} is obtained using the parameter ranges listed in Table~\ref{biao1}. We classify the numerical results in the $Y_{PU}$ and $Y_{XU}$ plane using
$\textcolor{blue}{\blacktriangle}~({\Gamma}_{NP}(h\rightarrow Z \gamma)/{\Gamma}_{SM}(h\rightarrow Z \gamma)<1.294),~\textcolor{orange}{\blacksquare}~(1.294\leq {\Gamma}_{NP}(h\rightarrow Z \gamma)/{\Gamma}_{SM}(h\rightarrow Z \gamma)<1.3),~ \textcolor{green}{\blacklozenge}~(1.3\leq {\Gamma}_{NP}(h\rightarrow Z \gamma)/{\Gamma}_{SM}(h\rightarrow Z \gamma)<1.35)$ and $\textcolor{red}{\bullet}~(1.35\leq {\Gamma}_{NP}(h\rightarrow Z \gamma)/{\Gamma}_{SM}(h\rightarrow Z \gamma)$. Here, $Y_{XU}$ and $Y_{PU}$ denote the Yukawa couplings between the SM-like and vector-like up-type quarks. Specifically, $Y_{XU}$ couples the SM right-handed quark $u_R$ to the vector-like left-handed quark $u_{XL}$, whereas $Y_{PU}$ represents the coupling between the left-hand and right-handed components of the vector-like quarks $u_{XL}$ and $u_{XR}$. As shown in Fig.~\ref{htozgamma1}, a substantial portion of the parameter space leads to a significant deviation of $\Gamma_{NP}(h\to Z\gamma)/\Gamma_{SM}(h\to Z\gamma)$  from the SM prediction. In particular, when $Y_{PU}$ is small or $Y_{XU}$ is in the moderate to high range, the red and green regions become dominant, indicating that the ratio can be enhanced to 1.30-1.35 or even higher. This corresponds to a deviation of about 30-35$\%$, far exceeding the theoretical uncertainties within the SM. In contrast, the regions close to the SM prediction (blue points) are mainly confined to the lower right corner of the plane, where $Y_{PU}$ is relatively large and $Y_{XU}$ remains small.

\begin{table*}
\caption{Scanning parameters for Fig.{\ref {htozgamma1}}}\label{biao1}
\begin{tabular}{|c|c|c|}
\hline
Parameters&Min&Max\\
\hline
$\hspace{1.5cm}g_X\hspace{1.5cm}$ &$\hspace{1.5cm}0.05\hspace{1.5cm}$&
$\hspace{1.5cm}0.85\hspace{1.5cm}$\\
\hline
$\hspace{1.5cm}g_{YX}\hspace{1.5cm}$ &$\hspace{1.5cm}-0.7\hspace{1.5cm}$& $\hspace{1.5cm}0.7\hspace{1.5cm}$\\
\hline
$\hspace{1.5cm}Y_{PU}\hspace{1.5cm}$ &$\hspace{1.5cm}0.055\hspace{1.5cm}$ &$\hspace{1.5cm}1\hspace{1.5cm}$\\
\hline
$\hspace{1.5cm}Y_{XU}\hspace{1.5cm}$ &$\hspace{1.5cm}0\hspace{1.5cm}$ &$\hspace{1.5cm}1\hspace{1.5cm}$\\
\hline
\end{tabular}
\end{table*}

Fixing the parameters at $Y_{PU}$=0.075 and $Y_{XU}$=1, we further plot in Fig.~\ref{htozgamma2} the dependence of ${\Gamma}_{NP}(h\rightarrow Z \gamma)/{\Gamma}_{SM}(h\rightarrow Z \gamma)$ on $g_X$. Here $g_X$ denotes the gauge coupling constant of the $U(1)_X$ group, while $g_{YX}$ represents the gauge kinetic mixing between $U(1)_Y$ and $U(1)_X$. As shown in Fig.~\ref{htozgamma2}, the ratio ${\Gamma}_{NP}(h\rightarrow Z \gamma)/{\Gamma}_{SM}(h\rightarrow Z \gamma)$ is nearly independent of $g_X$ and remains around 1.65 for all considered values of $g_{YX}$, corresponding to an enhancement of approximately 65$\%$ over the SM prediction.

\subsection{ The processes $h\rightarrow m_V Z$}
In this section, we analyze the decay processes $h\rightarrow m_V Z$. The decay constants of the vector mesons $\omega,\ \rho,\ J/\psi,\ \phi,\ \Upsilon$ are listed in Table~\ref{t1}.

\subsubsection{The process $h\rightarrow \omega Z$}
At first, we study the decay $h\rightarrow \omega Z$ and some suppositions are taken as $Y_{XE}=0.5,\ Y_{PE}=0.01,\ v_P=4500~\mathrm{GeV},\ v_S=8500~\mathrm{GeV}$.

\begin{figure}[ht]
\setlength{\unitlength}{5mm}
\centering
\includegraphics[width=2.9in]{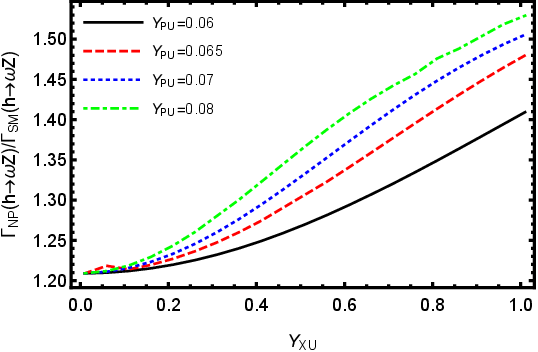}
\caption{${\Gamma}_{NP}(h\rightarrow \omega Z )/{\Gamma}_{SM}(h\rightarrow \omega Z)$ varying with $Y_{XU}$ for different $Y_{PU}$.}{\label {htoomegaz1}}
\end{figure}

For the parameter choice $g_X=0.6,\ g_{YX}=-0.1,\ Y_{XD}=0.8,\ Y_{PD}=0.01$, ${\Gamma}_{NP}(h\rightarrow \omega Z)/{\Gamma}_{SM}(h\rightarrow \omega Z)$ versus $Y_{XU}$ is shown in Fig.~\ref{htoomegaz1}. As can be seen, all four curves exhibit a monotonic increase from left to right, with the larger value of $Y_{PU}$ (the green line corresponding to $Y_{PU}$=0.08) lying above the others throughout the entire parameter range. The ratio always exceeds 1.2 across all parameter variations, and it can reach 1.45-1.50 around $Y_{XU}\approx 1$, corresponding to a sizable enhancement of about 20-50$\%$ relative to the SM prediction.

\begin{figure}[ht]
\setlength{\unitlength}{5mm}
\centering
\includegraphics[width=2.9in]{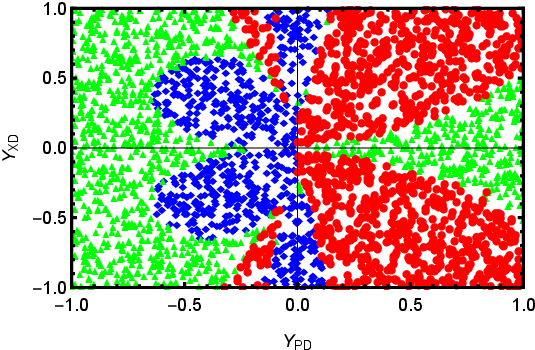}
\caption{${\Gamma}_{NP}(h\rightarrow \omega Z )/{\Gamma}_{SM}(h\rightarrow \omega Z )$ in $Y_{PD}-Y_{XD}$ plane, and the marking of the scattering points represents: $\textcolor{blue}{\blacklozenge}~({\Gamma}_{NP}(h\rightarrow \omega Z)/{\Gamma}_{SM}(h\rightarrow \omega Z )<1.58),~ \textcolor{green}{\blacktriangle}~(1.58\leq {\Gamma}_{NP}(h\rightarrow \omega Z )/{\Gamma}_{SM}(h\rightarrow \omega Z)<1.6),~ \textcolor{red}{\bullet}~(1.6\leq {\Gamma}_{NP}(h\rightarrow \omega Z)/{\Gamma}_{SM}(h\rightarrow \omega Z$).}{\label {htoomegaz2}}
\end{figure}

To further investigate the decay $h\rightarrow \omega Z$ and identify the regions of viable parameter space, we study the effects of the parameters $g_X,\ g_{YX},\ Y_{PD},\ Y_{XD}$. We generate a scatter plot in the ($Y_{PD}$,~$Y_{XD}$) plane under the condition $Y_{XU}=1, \ Y_{PU} = 0.075$. Using the parameter ranges given in Table~\ref{biao2}, the resulting distribution is shown in Fig.~\ref{htoomegaz2}. The scatter points are categorized as
$\textcolor{blue}{\blacklozenge}$ for ${\Gamma}_{NP}(h\rightarrow \omega Z)/{\Gamma}_{SM}(h\rightarrow \omega Z )<1.58$, $\textcolor{green}{\blacktriangle}$ for $1.58\leq {\Gamma}_{NP}(h\rightarrow \omega Z )/{\Gamma}_{SM}(h\rightarrow \omega Z)<1.6$ and $\textcolor{red}{\bullet}$ for $1.6\leq {\Gamma}_{NP}(h\rightarrow \omega Z)/{\Gamma}_{SM}(h\rightarrow \omega Z)$. The parameter $Y_{XD}$ controls the mixing between the vector-like down-type quarks and the third-generation SM quarks, while $Y_{PD}$ mainly determines the mass of the vector-like down-type quarks. It is evident that the parameter space exhibits an approximately symmetric distribution with respect to the horizontal axis. The blue region with a smaller ratio ($<1.58$) is confined to a narrow area slightly left of the center of the plot. Surrounding it is the green region with moderately larger values (1.58-1.60). Most notably, the vast majority of the parameter space, particularly regions with $Y_{PD}>0$ or $|Y_{XD}|>0.3$, is entirely dominated by red scatter points, corresponding to ${\Gamma}_{NP}(h\rightarrow \omega Z )/{\Gamma}_{SM}(h\rightarrow \omega Z)\ge 1.6$. This indicates that in these regions the decay width exceeds the SM prediction by at least 60$\%$.

\begin{table*}
\caption{Scanning parameters for Fig.{\ref {htoomegaz2}}}\label{biao2}
\begin{tabular}{|c|c|c|}
\hline
Parameters&Min&Max\\
\hline
$\hspace{1.5cm}g_X\hspace{1.5cm}$ &$\hspace{1.5cm}0.05\hspace{1.5cm}$&
$\hspace{1.5cm}0.85\hspace{1.5cm}$\\
\hline
$\hspace{1.5cm}g_{YX}\hspace{1.5cm}$ &$\hspace{1.5cm}-0.7\hspace{1.5cm}$& $\hspace{1.5cm}0.7\hspace{1.5cm}$\\
\hline
$\hspace{1.5cm}Y_{PD}\hspace{1.5cm}$ &$\hspace{1.5cm}-1\hspace{1.5cm}$ &$\hspace{1.5cm}1\hspace{1.5cm}$\\
\hline
$\hspace{1.5cm}Y_{XD}\hspace{1.5cm}$ &$\hspace{1.5cm}-1\hspace{1.5cm}$ &$\hspace{1.5cm}1\hspace{1.5cm}$\\
\hline
\end{tabular}
\end{table*}

\subsubsection{The process $h\rightarrow \rho Z$}
Secondly, we analyze the numerical results for the decay process $h\rightarrow \rho Z$ and set the parameters $g_X=0.6,\ g_{YX}=-0.1,\ Y_{PU}=0.075,\ Y_{PE}=0.01,\ v_P=4500~\mathrm{GeV},\ v_S=8500~\mathrm{GeV}$.

\begin{figure}[ht]
\setlength{\unitlength}{5mm}
\centering
\includegraphics[width=2.9in]{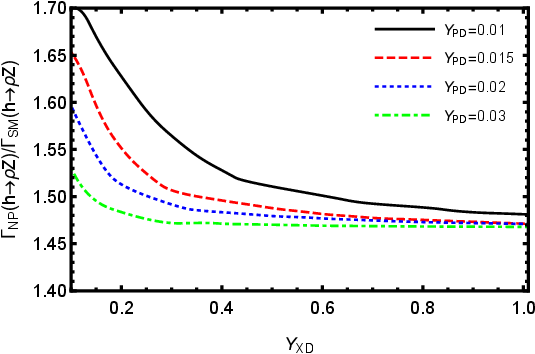}
\caption{${\Gamma}_{NP}(h\rightarrow \rho Z )/{\Gamma}_{SM}(h\rightarrow \rho Z)$ varying with $Y_{XD}$ for different $Y_{PD}$.}{\label {htorhoz1}}
\end{figure}

Assuming $\ Y_{XU}=1,\ Y_{XE}=0.5$, the variation of the ratio ${\Gamma}_{NP}(h\rightarrow \rho Z )/{\Gamma}_{SM}(h\rightarrow \rho Z)$  with $Y_{XD}$ is presented in Fig.~\ref{htorhoz1}, where the black, red, blue and green curves correspond to $Y_{PD}=0.01,\ 0.015,\ 0.02,\ 0.03$, respectively. Over the entire scanning range $Y_{XD}\in[0.1,\,1.0]$, the four curves lie significantly above unity, resulting in a deviation of approximately 45-70$\%$ from the SM prediction. All curves exhibit a peak in the region of small $Y_{XD}$, with the $Y_{PD}$=0.01 curve reaching the largest value, around 1.70. As $Y_{XD}$ increases, the curves show a decreasing trend and gradually approach a stable behavior, eventually converging to the interval of about 1.48-1.50 for $Y_{XD}\gtrsim 0.6$. Moreover, smaller values of $Y_{PD}$ correspond to a higher overall curve.

\begin{figure}[ht]
\setlength{\unitlength}{5mm}
\centering
\includegraphics[width=2.9in]{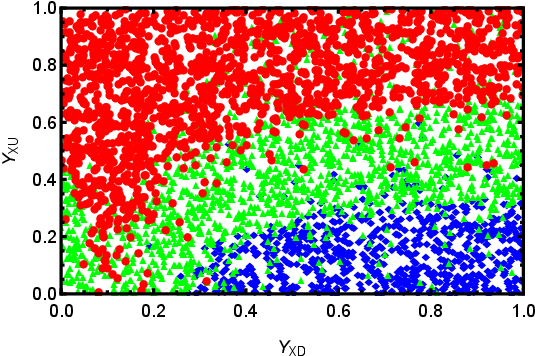}
\caption{${\Gamma}_{NP}(h\rightarrow \rho Z )/{\Gamma}_{SM}(h\rightarrow \rho Z )$ in $Y_{XD}-Y_{XU}$ plane, and the marking of the scattering points represents: $\textcolor{blue}{\blacklozenge}~({\Gamma}_{NP}(h\rightarrow \rho Z)/{\Gamma}_{SM}(h\rightarrow \rho Z )<1.26),~ \textcolor{green}{\blacktriangle}~(1.26\leq {\Gamma}_{NP}(h\rightarrow \rho Z )/{\Gamma}_{SM}(h\rightarrow \rho Z)<1.4),~ \textcolor{red}{\bullet}~(1.4\leq {\Gamma}_{NP}(h\rightarrow \rho Z)/{\Gamma}_{SM}(h\rightarrow \rho Z$).}{\label {htorhoz2}}
\end{figure}

To further explore the multidimensional behavior of the sensitive parameters, we fix $Y_{PD}=0.01$ and generate Fig. \ref{htorhoz2} by sampling points according to Table \ref{biao3}, illustrating the behavior of the $h\rightarrow \rho Z$ decay rate as a function of $Y_{XD}$ and $Y_{XU}$. The parameter plane is clearly divided into three regions: $\textcolor{blue}{\blacklozenge}$ correspond to cases where ${\Gamma}_{NP}(h\rightarrow \rho Z)/{\Gamma}_{SM}(h\rightarrow \rho Z ) < 1.26$, predominantly appearing in the low $Y_{XU}$ region ($Y_{XU}\lesssim 0.3$) and gradually extending toward larger $Y_{XD}$. $\textcolor{green}{\blacktriangle}$ represent ratios in the range 1.26-1.4, with a substantially broader distribution that roughly covers the intermediate region $Y_{XU}\sim 0.2-0.6$. $\textcolor{red}{\bullet}$ denote points for which the ratio exceeds 1.4, occupying almost the entire upper portion of the plane with large $Y_{XU}$ values ($Y_{XU}\gtrsim 0.6$), and they appear densely throughout the full interval $Y_{XD}\in[0,1]$. Overall, large values of $Y_{XU}$ yield ratios stably above 1.4, corresponding to an enhancement exceeding 40$\%$ compared with the SM prediction.

\begin{table*}
\caption{Scanning parameters for Fig.{\ref {htorhoz2}}}\label{biao3}
\begin{tabular}{|c|c|c|}
\hline
Parameters&Min&Max\\
\hline
$\hspace{1.5cm}Y_{XD}\hspace{1.5cm}$ &$\hspace{1.5cm}0\hspace{1.5cm}$&
$\hspace{1.5cm}1\hspace{1.5cm}$\\
\hline
$\hspace{1.5cm}Y_{XU}\hspace{1.5cm}$ &$\hspace{1.5cm}0\hspace{1.5cm}$& $\hspace{1.5cm}1\hspace{1.5cm}$\\
\hline
$\hspace{1.5cm}Y_{XE}\hspace{1.5cm}$ &$\hspace{1.5cm}0\hspace{1.5cm}$ &$\hspace{1.5cm}1\hspace{1.5cm}$\\
\hline
\end{tabular}
\end{table*}

\subsubsection{The process $h\rightarrow J/\psi Z$}
Thirdly, we perform a numerical analysis of the decay $h\rightarrow J/\psi Z$. The parameters are chosen as $Y_{XD}=0.8,\ Y_{PD}=0.01,\ Y_{XU}=1,\ v_P=4500~\mathrm{GeV}$.

\begin{figure}[ht]
\setlength{\unitlength}{5mm}
\centering
\includegraphics[width=2.9in]{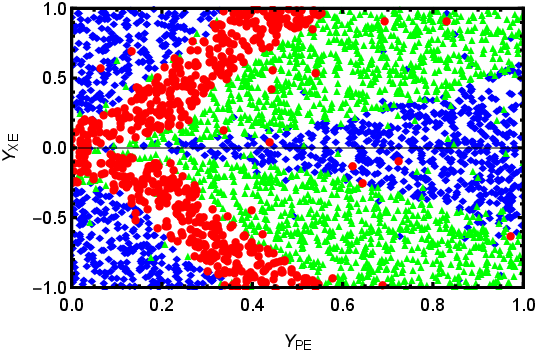}
\caption{${\Gamma}_{NP}(h\rightarrow J/\psi Z )/{\Gamma}_{SM}(h\rightarrow J/\psi Z )$ in $Y_{PE}-Y_{XE}$ plane, and the marking of the scattering points represents: $\textcolor{blue}{\blacklozenge}~({\Gamma}_{NP}(h\rightarrow J/\psi Z)/{\Gamma}_{SM}(h\rightarrow J/\psi Z )<1.245),~ \textcolor{green}{\blacktriangle}~(1.245\leq {\Gamma}_{NP}(h\rightarrow J/\psi Z )/{\Gamma}_{SM}(h\rightarrow J/\psi Z)<1.255),~ \textcolor{red}{\bullet}~(1.255\leq {\Gamma}_{NP}(h\rightarrow J/\psi Z)/{\Gamma}_{SM}(h\rightarrow J/\psi Z$).}{\label {htojz1}}
\end{figure}

We suppose the parameters with $Y_{PU}=0.075,\ v_S=8500~\mathrm{GeV}$. By sampling the parameter ranges listed in Table~\ref{biao4}, we obtain Fig.~\ref{htojz1}. In Fig.~\ref{htojz1}, the $\textcolor{blue}{\blacklozenge}$, $\textcolor{green}{\blacktriangle}$ and $\textcolor{red}{\bullet}$ correspond to ${\Gamma}_{NP}(h\rightarrow J/\psi Z)/{\Gamma}_{SM}(h\rightarrow J/\psi Z )<1.245$, $1.245\leq {\Gamma}_{NP}(h\rightarrow J/\psi Z )/{\Gamma}_{SM}(h\rightarrow J/\psi Z)<1.255$ and ${\Gamma}_{NP}(h\rightarrow J/\psi Z)/{\Gamma}_{SM}(h\rightarrow J/\psi Z)\geq1.255$, respectively. We examine the roles of $Y_{PE}$ and $Y_{XE}$ in Fig. \ref{htojz1}. In the lepton sector, the parameters $Y_{XE}$ and $Y_{PE}$ describe the Yukawa interactions that induce couplings between SM leptons and the vector-like fermions. The points exhibit symmetry about the horizontal axis $Y_{XE}=0$. As $Y_{PE}$ increases from 0 to 1, ${\Gamma}_{NP}(h\rightarrow J/\psi Z)/{\Gamma}_{SM}(h\rightarrow J/\psi Z )$ first increases and then decreases. Moreover, increasing $|Y_{XE}|$ enhances the ratio symmetrically across the parameter space when $Y_{PE}>0.2$, indicating that both positive and negative values of $Y_{XE}$ lead to similar enhancement effects. $\textcolor{red}{\bullet}$ show ratios consistently exceed 1.255, representing a clear deviation from the SM prediction.

\begin{table*}
\caption{Scanning parameters for Fig.{\ref {htojz1}}}\label{biao4}
\begin{tabular}{|c|c|c|}
\hline
Parameters&Min&Max\\
\hline
$\hspace{1.5cm}g_{X}\hspace{1.5cm}$ &$\hspace{1.5cm}0.05\hspace{1.5cm}$&
$\hspace{1.5cm}0.85\hspace{1.5cm}$\\
\hline
$\hspace{1.5cm}g_{YX}\hspace{1.5cm}$ &$\hspace{1.5cm}-0.7\hspace{1.5cm}$& $\hspace{1.5cm}0.7\hspace{1.5cm}$\\
\hline
$\hspace{1.5cm}Y_{PE}\hspace{1.5cm}$ &$\hspace{1.5cm}0\hspace{1.5cm}$ &$\hspace{1.5cm}1\hspace{1.5cm}$\\
\hline
$\hspace{1.5cm}Y_{XE}\hspace{1.5cm}$ &$\hspace{1.5cm}-1\hspace{1.5cm}$ &$\hspace{1.5cm}1\hspace{1.5cm}$\\
\hline
\end{tabular}
\end{table*}

\begin{figure}[ht]
\setlength{\unitlength}{5mm}
\centering
\includegraphics[width=2.9in]{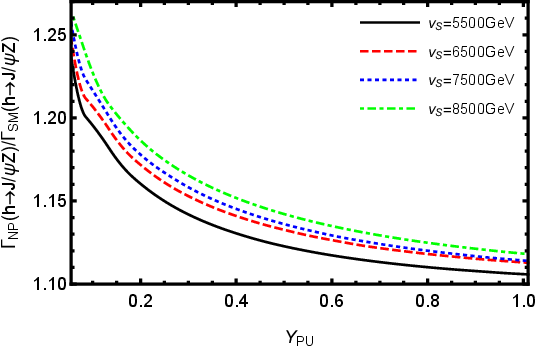}
\caption{${\Gamma}_{NP}(h\rightarrow J/\psi Z )/{\Gamma}_{SM}(h\rightarrow J/\psi Z)$ varying with $Y_{PU}$ for different $v_{S}$.}{\label {htojz2}}
\end{figure}

For the parameter set $g_{X}=0.6,\ g_{YX}=-0.1,\ Y_{XE}=0.5,\ Y_{PE}=0.01$, the variation of ${\Gamma}_{NP}(h\rightarrow J/\psi Z )/{\Gamma}_{SM}(h\rightarrow J/\psi Z)$ with respect to $Y_{PU}$ is shown in Fig.~\ref{htojz2}. The black ($v_S=5500\mathrm{GeV}$), red ($v_S=6500~\mathrm{GeV}$), blue ($v_S=7500~\mathrm{GeV}$) and green ($v_S=8500~\mathrm{GeV}$) curves correspond to different choices of the singlet scalar VEV $v_S$. As shown in Fig.~\ref{htojz2}, all four curves exhibit a monotonically decreasing behavior as $Y_{PU}$ increases, with the most pronounced deviations occurring in the small $Y_{PU}$ region. In particular, for $Y_{PU}\lesssim 0.1$, the ratio ${\Gamma}_{NP}(h\rightarrow J/\psi Z )/{\Gamma}_{SM}(h\rightarrow J/\psi Z)$ reaches 1.23-1.26, corresponding to an enhancement of more than 20$\%$ over the SM prediction. As $Y_{PU}$ continues to grow, the curves gradually flatten out. However, throughout the entire parameter range, the ratio remains above 1.10, still significantly larger than the SM value. Moreover, increasing $v_S$ raises the overall height of the curves, indicating that a larger singlet scalar VEV further strengthens the NP contributions to the $h\rightarrow J/\psi Z$ decay.

\subsubsection{The process $h\rightarrow \phi Z$}
Then, we analyze the decay process $h\rightarrow \phi Z$ numerically with $g_{X}=0.6,\ g_{YX}=-0.1,\ Y_{XD}=0.8,\ Y_{XU}=1,\ Y_{XE}=0.5$.

\begin{figure}[ht]
\setlength{\unitlength}{5mm}
\centering
\includegraphics[width=2.9in]{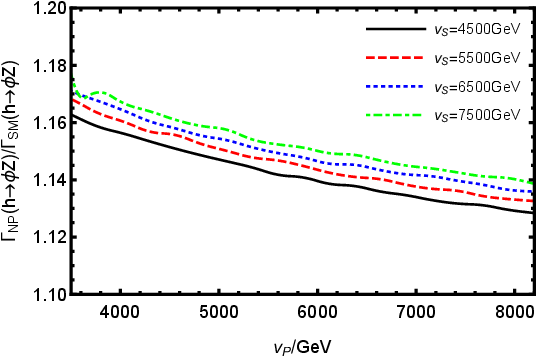}
\caption{${\Gamma}_{NP}(h\rightarrow \phi Z )/{\Gamma}_{SM}(h\rightarrow \phi Z)$ varying with $v_P$ for different $v_S$.}{\label {htophiz1}}
\end{figure}

In Fig.~\ref{htophiz1}, we fix the parameters as $Y_{PD}=0.01,\ Y_{PU}=0.075,\ Y_{PE}=0.01$ and plot the ratio ${\Gamma}_{NP}(h\rightarrow \phi Z )/{\Gamma}_{SM}(h\rightarrow \phi Z)$ as a function of $v_P$, where $v_P$ denotes the VEV of the singlet scalar $P$. From the Fig.~\ref{htophiz1}, one observes that all four curves exhibit a decreasing trend. The green curve ($v_S=7500\mathrm{GeV}$) lies above the blue one ($v_S=6500~\mathrm{GeV}$), which in turn lies above the red ($v_S=5500~\mathrm{GeV}$), followed by the black curve ($v_S=4500~\mathrm{GeV}$) at the lowest position. As $v_P$ increases, the ratio ${\Gamma}_{NP}(h\rightarrow \phi Z )/{\Gamma}_{SM}(h\rightarrow \phi Z)$ decreases, whereas increasing $v_S$ has the opposite effect and enhances the ratio. The maximal value of the ratio reaches about 1.18, corresponding to a deviation of roughly 20$\%$ from the SM prediction.

\begin{figure}[ht]
\setlength{\unitlength}{5mm}
\centering
\includegraphics[width=2.9in]{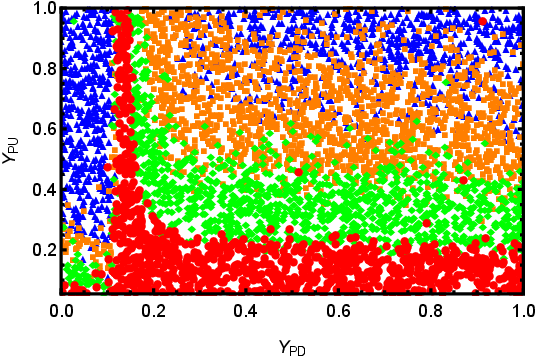}
\caption{${\Gamma}_{NP}(h\rightarrow \phi Z )/{\Gamma}_{SM}(h\rightarrow \phi Z )$ in $Y_{PD}-Y_{PU}$ plane, and the marking of the scattering points represents: $\textcolor{blue}{\blacktriangle}~({\Gamma}_{NP}(h\rightarrow \phi Z)/{\Gamma}_{SM}(h\rightarrow \phi Z)<1.15),~\textcolor{orange}{\blacksquare}~(1.15\leq {\Gamma}_{NP}(h\rightarrow \phi Z)/{\Gamma}_{SM}(h\rightarrow \phi Z)<1.17),~ \textcolor{green}{\blacklozenge}~(1.17\leq {\Gamma}_{NP}(h\rightarrow \phi Z)/{\Gamma}_{SM}(h\rightarrow \phi Z)<1.2),~ \textcolor{red}{\bullet}~(1.2\leq {\Gamma}_{NP}(h\rightarrow \phi Z)/{\Gamma}_{SM}(h\rightarrow \phi Z$).}{\label {htophiz2}}
\end{figure}

Next, we perform a scatter analysis of the $h\rightarrow \phi Z$ process using the parameter ranges listed in Table~\ref{biao5}. For $v_P=4500~\mathrm{GeV}$ and $v_S=8500~\mathrm{GeV}$, the distribution of points in the ($Y_{PD},\,Y_{PU}$) plane is shown in Fig.~\ref{htophiz2}. The red $\textcolor{red}{\bullet}$ (corresponding to ratios larger than 1.2) are mainly concentrated in the low $Y_{PU}$ region with $Y_{PU}\lesssim 0.25$, indicating that the NP effects are most pronounced for small $Y_{PU}$, where the decay width can be enhanced by more than 20$\%$ relative to the SM prediction. As $Y_{PU}$ increases, $\textcolor{red}{\bullet}$ gradually transition into $\textcolor{green}{\blacklozenge}$ and $\textcolor{orange}{\blacksquare}$, which correspond to intermediate ranges of 1.17-1.20 and 1.15-1.17, respectively. This behavior suggests that the NP contribution becomes weaker but still leads to deviations at the level of roughly 15$\%$. Finally, in the region of larger $Y_{PU}$, the $\textcolor{blue}{\blacktriangle}$ dominate, corresponding to ratios below 1.15, where the NP corrections are close to yet remain noticeably above the SM prediction.

\begin{table*}
\caption{Scanning parameters for Fig.{\ref {htophiz2}}}\label{biao5}
\begin{tabular}{|c|c|c|}
\hline
Parameters&Min&Max\\
\hline
$\hspace{1.5cm}Y_{PD}\hspace{1.5cm}$ &$\hspace{1.5cm}0\hspace{1.5cm}$& $\hspace{1.5cm}1\hspace{1.5cm}$\\
\hline
$\hspace{1.5cm}Y_{PU}\hspace{1.5cm}$ &$\hspace{1.5cm}0.055\hspace{1.5cm}$ &$\hspace{1.5cm}1\hspace{1.5cm}$\\
\hline
$\hspace{1.5cm}Y_{PE}\hspace{1.5cm}$ &$\hspace{1.5cm}0\hspace{1.5cm}$ &$\hspace{1.5cm}1\hspace{1.5cm}$\\
\hline
\end{tabular}
\end{table*}

\subsubsection{The process $h\rightarrow \Upsilon Z$}
At last, we carry out a numerical analysis of the decay $h\rightarrow \Upsilon Z$. As a heavy vector meson, $\Upsilon$ is composed of a $b\bar b$ pair.

\begin{figure}[ht]
\setlength{\unitlength}{5mm}
\centering
\includegraphics[width=2.9in]{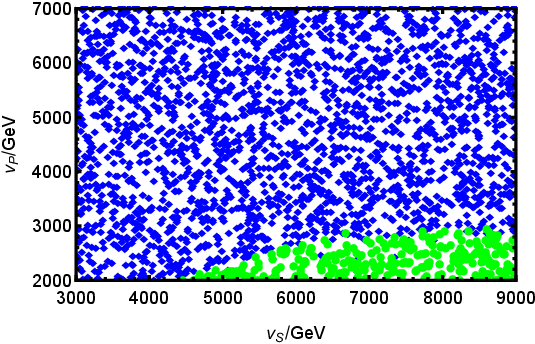}
\caption{${\Gamma}_{NP}(h\rightarrow \Upsilon Z )/{\Gamma}_{SM}(h\rightarrow \Upsilon Z )$ in $v_S-v_P$ plane, and the marking of the scattering points represents: $\textcolor{blue}{\blacklozenge}~({\Gamma}_{NP}(h\rightarrow \Upsilon Z)/{\Gamma}_{SM}(h\rightarrow \Upsilon Z )<1.01),~ \textcolor{green}{\bullet}~(1.01\leq {\Gamma}_{NP}(h\rightarrow \Upsilon Z )/{\Gamma}_{SM}(h\rightarrow \Upsilon Z)$).}{\label {htouz1}}
\end{figure}

With the parameter choices $g_X = 0.6$, $g_{YX} = -0.1$, $Y_{XD} = 0.8$, $Y_{PD} = 0.01$, $Y_{XU} = 1$, $Y_{PU} = 0.075$, $Y_{XE} = 0.5$, $Y_{PE} = 0.01$, we investigate the impact of $v_S$ and $v_P$ on the ratio $\Gamma_{\mathrm{NP}}(h \to \Upsilon Z)/\Gamma_{\mathrm{SM}}(h \to \Upsilon Z)$. The results are shown in Fig.~\ref{htouz1}, based on the scan ranges provided in Table~\ref{biao6}. The $\textcolor{blue}{\blacklozenge}$ and $\textcolor{green}{\bullet}$ denote regions with ${\Gamma}_{NP}(h\rightarrow \Upsilon Z)/{\Gamma}_{SM}(h\rightarrow \Upsilon Z )<1.01$ and $1.01\leq {\Gamma}_{NP}(h\rightarrow \Upsilon Z )/{\Gamma}_{SM}(h\rightarrow \Upsilon Z)$, respectively. It can be observed that the larger corrections ($\textcolor{green}{\bullet}$) occur predominantly in the lower right region of the parameter plane. This indicates that larger values of $v_S$ combined with smaller values of $v_P$ enhance the NP contributions. The ratio reaches values in the range of approximately 1.01-1.35. For the decay $h\rightarrow \Upsilon Z$, the magnitude of the NP effects is smaller than in the other processes considered.

\begin{table*}
\caption{Scanning parameters for Fig.{\ref {htouz1}}}\label{biao6}
\begin{tabular}{|c|c|c|}
\hline
Parameters&Min&Max\\
\hline
$\hspace{1.5cm}v_{S}/\rm GeV\hspace{1.5cm}$ &$\hspace{1.5cm}3000\hspace{1.5cm}$&
$\hspace{1.5cm}9000\hspace{1.5cm}$\\
\hline
$\hspace{1.5cm}v_{P}/\rm GeV\hspace{1.5cm}$ &$\hspace{1.5cm}2000\hspace{1.5cm}$& $\hspace{1.5cm}7000\hspace{1.5cm}$\\
\hline
\end{tabular}
\end{table*}

\section{discussion and conclusion}\label{sec5}
In summary, the $U(1)_X$VLFM model introduces an additional Abelian gauge symmetry and one generation of vectorlike fermions, leading to a noticeable modification of the Higgs gauge interaction structure compared with the SM. The right-handed neutrinos and the two singlet Higgs fields included in the model can realize the seesaw mechanism within a non-supersymmetric framework, while the extended fermion spectrum also provides new contributions to Higgs related loop effects. Within this framework, we perform a detailed analysis of the rare decay $h\rightarrow Z\gamma$ as well as the hadronic channels $h\rightarrow m_V Z$ with $m_V=\rho,\omega,\phi,J/\psi,\Upsilon$. For completeness, numerical results for $h\to\gamma\gamma$ and $h\to V V^* (V = Z, W)$ are also evaluated to assess the broader impact of NP on Higgs decay patterns. In the SM, the $hZZ$ coupling appears at tree level, whereas the $h\gamma Z$ coupling arises only through loop contributions. For the decay $h\to m_V Z$, the amplitude receives both direct and indirect contributions. The direct contribution corresponds to the Higgs coupling to quarks that hadronize directly into the final state vector meson, while the indirect contribution originates from the Higgs coupling to an on-shell $Z$ boson and an off-shell gauge boson ($\gamma$ or $Z$), which subsequently converts into the vector meson. As pointed out in Ref.~\cite{htomz}, the indirect contribution is typically much larger than the direct one in the viable parameter space. Beyond the SM, the $h\gamma Z$ vertex can receive new CP-even and CP-odd effective couplings $C_{\gamma Z}$ and $\tilde C_{\gamma Z}$, although the CP-even part usually remains dominant. In this work, we compute their contributions to the $h\gamma Z$ vertex using the effective Lagrangian approach.

In the SM, the Higgs boson couples to the $W$ and $Z$ gauge bosons at tree level, while its interaction with photons arises only through loop diagrams. Consequently, the decay channels $h\to\gamma\gamma$, $h\to WW^*$ and $h\to ZZ^*$ exhibit high sensitivity in experimental measurements, with the observed signal strengths $R_{\gamma\gamma}=1.10\pm0.06,\ R_{WW^*}=1.00\pm0.08,\ R_{ZZ^*}=1.02\pm0.08$~\cite{ATLAS:2016neq,gamma1,gamma2,gamma4,zz}.
The dominant contributions to the processes $h\to Z Z^*,~h\to\gamma\gamma$ and $h\rightarrow Z\gamma$ all originate from similar one loop topologies. Within the $U(1)_X$VLFM framework examined in this work, the NP corrections to $h\to\gamma\gamma$ typically fall in the range 1.0-1.2, while those for $h\to VV^* (V=Z,W)$ are around 1.1. Unlike $h\to\gamma\gamma$ and $h\to Z\gamma$, which are purely loop induced and sensitive only to charged particles running in the loop, the decay $h\to ZZ^*$ also receives a tree level contribution from the $hZZ$ vertex and allows neutral particles to participate in the loop diagrams. Despite these differences, the three decay channels remain structurally very similar. It is worth emphasizing that the combined ATLAS and CMS analysis reports a signal strength of $\mu_{Z\gamma}=2.2\pm0.7$ for $h\to Z\gamma$, significantly higher than the SM expectation and reaching the level of experimental evidence~\cite{hZgexp}. This deviation further motivates the study of possible NP effects in $h\to Z\gamma$ and in the related $h\to m_V Z$ decay channels.

The numerical analysis shows that the NP correction to $h\rightarrow Z\gamma$ can reach up to 65$\%$ compared with the SM prediction. For the decays $h\rightarrow \omega Z$ and $h\rightarrow \rho Z$, the NP contributions lie in the range of 20$\%$-70$\%$, while for $h\rightarrow \phi Z$ and $h\rightarrow J/\psi Z$, the corrections are typically 10$\%$-25$\%$. Among the vector mesons considered in this study ($\rho,\ \omega,\ \phi,\ J/\psi,\ \Upsilon$), the $\Upsilon$ is the heaviest. Although we tried adjusting many parameters in an attempt to enhance the deviation in $h\rightarrow \Upsilon Z$, the NP contribution to this process remains rather small. Overall, our results suggest a clear trend: NP effects in $h\rightarrow m_V Z$ are more pronounced when the final state vector meson is light. Considering that ${\Gamma}_{NP}(h\rightarrow Z\gamma)/{\Gamma}_{SM}(h\rightarrow Z\gamma)\approx 1.65$, the fact that the NP corrections to the rare decays $h\rightarrow m_V Z$ can improve about 70$\%$ of the SM prediction represents a sizable deviation.

In NP scenarios, the additional interactions can modify the SM CP-even coupling and may also introduce CP-odd contributions, thereby enhancing the branching ratio of the rare decay $h\to m_V Z$. Since this decay is intrinsically rare, the current experimental sensitivity is still insufficient for detection. However, its branching ratio can be reliably calculated at the theoretical level, and it may become observable at the HL-LHC and future high-energy colliders. Therefore, the decay channels $h\to m_V Z$ provide an important opportunity to test the $U(1)_X$VLFM model and to explore possible NP effects in the Higgs sector.

\begin{acknowledgments}
This work is supported by National Natural Science Foundation of China (NNSFC)
(No.12075074), Natural Science Foundation of Hebei Province
(A2023201040, A2022201022, A2022201017, A2023201041), Natural Science Foundation of Hebei Education Department (QN2022173), the Project of the China Scholarship Council (CSC) No. 202408130113. X. Dong acknowledges support from Funda\c{c}\~{a}o para a Ci\^{e}ncia e a Tecnologia (FCT, Portugal) through the projects CFTP FCT Unit UIDB/00777/2020, UIDP/00777/2020 and UID/00777/2025.
\end{acknowledgments}

\appendix
\section{mass matrix and coupling in $U(1)_X$VLFM}\label{A1}
The mass matrix for up-type quark in the $(u_L, u_{XL})$, $(u^*_R, u^*_{XR})$ basis reads
\begin{equation}
m_u = \left(
\begin{array}{cc}
\frac{1}{\sqrt{2}}v Y^T_u &0\\
\frac{1}{\sqrt{2}}v_S Y^T_{XU}  &\frac{1}{\sqrt{2}}v_P Y^T_{PU}\end{array}
\right).
\end{equation}
We diagonalize the mass matrix using $U^u_L$ and $U^u_R$
\begin{equation}
U_L^{u,*}\, m_u \, U_R^{u,\dagger} = m_u^{dia}.
\end{equation}

In the $(e_L, e_{XL}), (e^*_R, e^*_{XR})$ basis, the lepton mass matrix is given by
\begin{equation}
m_e = \left(
\begin{array}{cc}
\frac{1}{\sqrt{2}}v Y^T_e &0\\
\frac{1}{\sqrt{2}}v_S Y^T_{XE}  &\frac{1}{\sqrt{2}}v_P Y^T_{PE}\end{array}
\right).
\end{equation}
This matrix is diagonalized by $U^e_L$ and $U^e_R$
\begin{equation}
U_L^{e,*}\, m_e \, U_R^{e,\dagger} = m_e^{dia}.
\end{equation}

Other Higgs-related vertices are included as follows
\begin{eqnarray}
&&\mathcal{L}_{h_k\bar e_i e_j}=\bar{e}_i\Big\{-i \frac{1}{\sqrt 2} \Big(\sum_{b=1}^3\sum_{a=1}^3U^{e,*}_{L,jb}U^{e,*}_{R,ia}Y_{e, a b}Z^H_{k1}+U^{e,*}_{L,j4}\sum_{a=1}^3U^{e,*}_{R,ia}Y_{XE,a1}Z^H_{k2}\nonumber\\
&&\hspace{1.6cm}+U^{e,*}_{L,j4}U^{e,*}_{R,i4}Y_{PE}Z^H_{k3}\Big)P_L\nonumber\\
&&\hspace{1.6cm}-i \frac{1}{\sqrt 2} \Big(\sum_{b=1}^3\sum_{a=1}^3U^{e}_{R,ja}U^{e}_{L,ib}Y^*_{e, a b}Z^H_{k1}+U^{e}_{L,i4}\sum_{a=1}^3U^{e}_{R,ja}Y^*_{XE,a1}Z^H_{k2}\nonumber\\
&&\hspace{1.6cm}+U^{e}_{R,j4}U^{e}_{L,i4}Y^*_{PE}Z^H_{k3}\Big)P_R\Big\}e_j h_k,
\end{eqnarray}
\begin{eqnarray}
&&\mathcal{L}_{h_k\bar d_i d_j}=\bar{d}_i\Big\{-i \frac{1}{\sqrt 2} \Big(\sum_{b=1}^3\sum_{a=1}^3U^{d,*}_{L,jb}U^{d,*}_{R,ia}Y_{d, a b}Z^H_{k1}+U^{d,*}_{L,j4}\sum_{a=1}^3U^{d,*}_{R,ia}Y_{XD,a1}Z^H_{k2}\nonumber\\
&&\hspace{1.6cm}+U^{d,*}_{L,j4}U^{d,*}_{R,i4}Y_{PD}Z^H_{k3}\Big)P_L\nonumber\\
&&\hspace{1.6cm}-i \frac{1}{\sqrt 2} \Big(\sum_{b=1}^3\sum_{a=1}^3U^{d}_{R,ja}U^{d}_{L,ib}Y^*_{d, a b}Z^H_{k1}+U^{d}_{L,i4}\sum_{a=1}^3U^{d}_{R,ja}Y^*_{XD,a1}Z^H_{k2}\nonumber\\
&&\hspace{1.6cm}+U^{d}_{R,j4}U^{d}_{L,i4}Y^*_{PD}Z^H_{k3}\Big)P_R\Big\}d_j h_k,
\end{eqnarray}
\begin{eqnarray}
&&\mathcal{L}_{h_k\bar u_i u_j}=\bar{u}_i\Big\{-i \frac{1}{\sqrt 2} \Big(\sum_{b=1}^3\sum_{a=1}^3U^{u,*}_{L,jb}U^{u,*}_{R,ia}Y_{u, a b}Z^H_{k1}+U^{u,*}_{L,j4}\sum_{a=1}^3U^{u,*}_{R,ia}Y_{XU,a1}Z^H_{k2}\nonumber\\
&&\hspace{1.6cm}+U^{u,*}_{L,j4}U^{u,*}_{R,i4}Y_{PU}Z^H_{k3}\Big)P_L\nonumber\\
&&\hspace{1.6cm}-i \frac{1}{\sqrt 2} \Big(\sum_{b=1}^3\sum_{a=1}^3U^{u}_{R,ja}U^{u}_{L,ib}Y^*_{u, a b}Z^H_{k1}+U^{u}_{L,i4}\sum_{a=1}^3U^{u}_{R,ja}Y^*_{XU,a1}Z^H_{k2}\nonumber\\
&&\hspace{1.6cm}+U^{u}_{R,j4}U^{u}_{L,i4}Y^*_{PU}Z^H_{k3}\Big)P_R\Big\}u_j h_k,
\end{eqnarray}
with
\begin{eqnarray}
&&Y_{XE,a 1} = \left(
\begin{array}{c}
0\\
0\\
Y_{XE}\end{array}
\right), \quad
Y_{XD,a 1} = \left(
\begin{array}{c}
0\\
0\\
Y_{XD}\end{array}
\right),\quad
Y_{XU,a 1} = \left(
\begin{array}{c}
0\\
0\\
Y_{XU}\end{array}
\right).
\end{eqnarray}

\end{document}